\documentclass[a4paper,10pt]{article}
\usepackage{jheppub}
\usepackage{booktabs}
\usepackage{array}
\usepackage{epsf,epsfig,subfigure,mathtools,hhline,float,multirow,nicefrac,slashed,color,url,graphicx,subfigure}
\usepackage[utf8x]{inputenc}
\usepackage[section]{placeins}
\usepackage[outdir=./]{epstopdf}
\newcommand{\be}{\begin{equation}}
\newcommand{\ee}{\end{equation}}
\newcommand{\bes}{\begin{equation*}}
\newcommand{\ees}{\end{equation*}}
\newcommand{\bea}{\begin{eqnarray}}
\newcommand{\eea}{\end{eqnarray}}
\newcommand{\beas}{\begin{eqnarray*}}
\newcommand{\eeas}{\end{eqnarray*}}

\title{On the CP Nature of the `95 GeV' Anomalies}
\author[a]{Tanmoy Mondal,}
\author[b,c]{Stefano Moretti}
\author[d,e]{and Prasenjit Sanyal}
\affiliation[a]{Birla Institute of Technology and Science, Pilani, 333031, Rajasthan, India}
\affiliation[b]{School of Physics and Astronomy, University of Southampton, Southampton SO17 1BJ, United Kingdom}
\affiliation[c]{Department of Physics and Astronomy, Uppsala University, Box 516, SE-751 20 Uppsala, Sweden}
\affiliation[d]{Department of Physics, Konkuk University, Seoul 05029, Republic of Korea}
\affiliation[e]{Center for Quantum Spacetime, Sogang University, 35 Baekbeom-ro, Seoul 121-742, Republic of Korea}

\emailAdd{tanmoy.mondal@pilani.bits-pilani.ac.in}
\emailAdd{s.moretti@soton.ac.uk}
\emailAdd{stefano.moretti@physics.uu.se}
\emailAdd{psanyal@sogang.ac.kr}

\abstract{Under the assumption that the various evidences of a `95 GeV' excess, seen in data at the Large Electron Positron (LEP) collider as well as the Large Hadron Collider (LHC), correspond to actual signals of new physics Beyond the Standard Model (BSM), we characterise the underlying particle explaining these anomalies in terms of its Charge/Parity (CP) quantum numbers. In doing so, we use $\chi^2$ fits to test the CP-even (scalar) and CP-odd (pseudoscalar) hypotheses and superpositions of these, thus under the assumption of a spin-0 resonance. This is done through the exploitation of $\tau^+\tau^-$ decays, in both their fully hadronic and semi-leptonic modes, in a 
model-independent way, so that our approach enables one to test a variety of BSM hypotheses, having proven here that the High-Luminosity LHC (HL-LHC) will be in a position  to disentangle the CP  nature of such a new particle 
within $\pm(0.27-0.47)$ radians of the true hypothesis at $90\%$ Confidence Level (CL), depending on the assumed background systematics.}

\preprint{CQUeST-2025-0763}
\date{}
\begin{document}

\maketitle  
\section{Introduction}

Since the discovery of the 125 GeV Higgs boson at the  Large Hadron Collider (LHC) in 2012~\cite{ATLAS:2012yve,CMS:2012qbp}, significant efforts have been made by the ATLAS and CMS collaborations in ascertaining the nature of this particle, that is, measuring its width, couplings, spin and CP quantum numbers. All corresponding measurements are pointing towards the discovered Higgs boson being the Standard Model (SM) one. Despite all this, though, the search for physics Beyond the SM (BSM) continues, spurred, on the theoretical side, by its many flaws and, on the experimental side,   by the ever increasing  precision of the analyses as well as additional data sets being collected during Run 3 of the LHC.

The 2-Higgs Doublet Model (2HDM)~\cite{Gunion:1992hs,Branco:2011iw}, as embedded in various BSM scenarios, including fundamental ones like Supersymmetry \cite{Moretti:2019ulc} and Compositeness \cite{DeCurtis:2018zvh,DeCurtis:2018iqd}, is particularly interesting, as it offers in its particle spectrum both a CP-even and a CP-odd (neutral) Higgs boson, alongside the discovered SM-like Higgs state,  both of which can in fact be lighter or heavier than the latter\footnote{The CP-even (neutral) Higgs states of the 2HDM are labelled as $h$ and $H$ (with $m_h<m_H$) while the CP-odd (neutral) one as $A$, plus there are also two charged Higgs bosons, $H^\pm$.}. Hence, this BSM scenario has been studied extensively since its  inception. However, its general version allows for non-diagonal Yukawa couplings, potentially leading to large  Flavor Changing Neutral Currents (FCNCs) at tree level, contrary to experimental evidence. To constrain the latter, from the theoretical side, a $Z_2$ symmetry (which can be softly broken, in fact)  is typically imposed to define the coupling structure of the two Higgs doublets to SM fermions. This classification includes the so-called Type-I, Type-II, lepton-specific and flipped scenarios \cite{Branco:2011iw}, alongside the 2HDM Type-III, which allows direct couplings of both doublets to all SM fermions. The 2HDM Yukawa structure is then refined by both theoretical consistency requirements and experimental measurements of the SM-like Higgs mass and couplings or limits on the same parameters for the other Higgs states. 

Given that a 2HDM, as mentioned, allows (in some Types) for neutral Higgs bosons states lighter than 125 GeV, both ATLAS and CMS have pursued this possibility through dedicated searches in presence of a mass configuration wherein $m_h$ and/or $m_A<m_H=125$ GeV (with $m_{H^\pm}$ also rather small, to comply with various high-precision data). (Hereafter, in our notation, when referring to the 2HDM, the CP-even $H$ state is the SM-like one already discovered while the $h$ and $A$ states are the lightest CP-even and CP-odd Higgs bosons, respectively.) However, in performing such searches, neither ATLAS nor CMS have defined potential signal regions based on 2HDM assumptions, as therein the latter only entered at the interpretation level. Therefore, their results can be used in a model-independent way, which is what we will be doing here by introducing a simplified model approach wherein the underlying BSM scenario is parameterized solely in terms of masses and (gauge plus Yukawa) couplings that are fitted to data, yet, we assume the same particle content of the generic 2HDM when computing the decay width of a new Higgs state that we will introduce (hereafter, denoted by $S$).  

In fact, during searches for a low mass neutral Higgs boson (i.e., our $S$ state), the CMS collaboration reported an excess near 95 GeV in the di-photon  invariant mass already back in 2018 \cite{CMS:2018cyk}. In March 2023, they  confirmed such an excess, eventually claiming a local significance of 2.9$\sigma$ computed at $m_{\gamma\gamma}=95.4$ GeV \cite{CMS:2022goy,CMS:2023yay}. Similarly, ATLAS observed an excess at about 95 GeV with a local significance of 1.7$\sigma$, aligning with CMS findings \cite{Arcangeletti, ATLAS:2023jzc}. Furthermore, the CMS collaboration has reported an excess in the search for a light neutral (pseudo)scalar Higgs boson decaying into $\tau^+\tau^-$ pairs, with local significance of $2.6\sigma$ around a mass of 95 GeV. (Yet, other CMS data tend to  invalidate attempts   to attribute such a di-tau excess to a new Higgs boson  because searches for a spin-0 resonance produced  in \(t\bar{t}\)-associated production and decaying into $\tau$ pairs seem not to support this interpretation \cite{CMS:2022arx}.) Finally, the Large Electron Positron (LEP) collider collaborations~\cite{LEPWorkingGroupforHiggsbosonsearches}
explored the low Higgs mass domain extensively in the $e^+e^-\to ZS$ production mode, with $S$ decaying via the 
$\tau^+\tau^-$ and $b\overline{b}$ channels. Interestingly for our purposes is the fact that an excess had  been 
reported in 2006
in the $e^+e^- \to Z  b\overline{b}$ channel  for  $m_{b\bar b}$ around 98 GeV~\cite{ALEPH:2006tnd}.
Given the limited mass resolution of the di-jet invariant mass at LEP, though, this anomaly may well coincide with the discussed  excesses seen by CMS and/or ATLAS in the $\gamma\gamma$ and $\tau^+\tau^-$ final states.

 Since all the excesses appear in very similar mass regions, several studies~\cite{Cao:2016uwt,Heinemeyer:2021msz,Biekotter:2021qbc,Biekotter:2019kde,Cao:2019ofo,Biekotter:2022abc,Iguro:2022dok,Li:2022etb,Cline:2019okt,Biekotter:2021ovi,Crivellin:2017upt,Cacciapaglia:2016tlr,Abdelalim:2020xfk,Biekotter:2022jyr,Biekotter:2023jld,Azevedo:2023zkg,Biekotter:2023oen,Cao:2024axg,Wang:2024bkg,Li:2023kbf,Dev:2023kzu,Borah:2023hqw,Cao:2023gkc,Ellwanger:2023zjc,Aguilar-Saavedra:2023tql,Ashanujjaman:2023etj,Bhattacharya:2023lmu,Dutta:2023cig,Ellwanger:2024txc,Diaz:2024yfu,Ellwanger:2024vvs,Ayazi:2024fmn, Arhrib:2024wjj, Gao:2024qag} have explored the possibility of simultaneously explaining these anomalies within BSM frameworks featuring a Higgs state lighter than 125 GeV, while being in agreement with current measurements of the properties of the SM-like Higgs state observed at the LHC. In the attempt to explain the excesses in the $\gamma\gamma$ and $b\overline{b}$ channels, it was found in Refs.~\cite{Benbrik:2022azi, Benbrik:2022tlg, Belyaev:2023xnv,Belyaev:2024lah,Benbrik:2024ptw} that the 2HDM Type-III with a particular Yukawa texture can
successfully accommodate both measurements simultaneously with the lightest CP-even Higgs boson of the model, while being consistent with all relevant theoretical and experimental constraints. In fact, also a 2HDM Type-I can afford similar explanations to the excesses \cite{Khanna:2024bah}.
Further recent studies have shown that actually all three aforementioned signatures can be simultaneously explained in the 2HDM plus a real (N2HDM)~\cite{Biekotter:2022jyr} or a complex (S2HDM) \cite{Biekotter:2023jld,Biekotter:2023oen} singlet.

The purpose of this paper is to assume that, indeed, a spin-0 particle of yet undefined CP is behind all such anomalies. However,
amongst the latter, we focus here on the $\tau^+\tau^-$ one, as this has some diagnostic power in terms of pinning down its  CP properties, as shown in Refs.~\cite{Bower:2002zx,Desch:2003rw,Berge:2013jra,Berge:2008wi,Berge:2008dr,Berge:2011ij,Berge:2014sra,Berge:2014wta,Berge:2015nua,Han:2016bvf,Antusch:2020ngh,Dutta:2021del,Esmail:2024jdg}. So, let us spend some time on this particular anomaly. 
The CMS collaboration has reported such an excess in the $\tau^+\tau^-$ final state across all $\tau$ decay modes (leptonic and hadronic) for both leptons. The local and global significances of this particular excess are  $2.6 \sigma$ and $2.3 \sigma$  for $m_{\tau^+\tau^-} = 95$ GeV, respectively. At $ m_{\tau^+\tau^-} = 100$ GeV, these values are $3.1\sigma$ and $2.7\sigma$. By introducing an additional single neutral narrow resonance $S$,
the best fit values to the signal rate for this excess are~\cite{CMS-PAS-HIG-21-001}:
\begin{eqnarray}\label{eq:excess_ta}
\sigma(gg \to S)\;{\rm BR}(S \to \tau^+\tau^-) = 7.7^{+3.9}_{-3.1} ~\text{pb} \quad \text{for}~ m_{S} = 95 ~\text{GeV},  
\label{95 tautau excess}\\
\sigma(gg \to S)\;{\rm BR}(S \to \tau^+\tau^-) = 5.8^{+2.4}_{-2.0} ~\text{pb} \quad \text{for}~ m_{S} = 100 ~\text{GeV}.
\end{eqnarray}
In our work, we assume a 95 GeV resonance. Interpreting Eq.~(\ref{95 tautau excess}) in terms of a signal strength, we get 
\begin{eqnarray}\label{eq:excess_tautau}
\mu^{\text{exp}}_{\tau^+\tau^-} = \frac{\sigma^{\text{exp}} (gg \to S \to \tau^+\tau^-)}{\sigma^{\text{SM}} (gg \to \Phi \to \tau^+\tau^-)} = 1.2 \pm 0.5,
\end{eqnarray}
where we use a symmetric uncertainty interval for such a signal strength, which is obtained from
the lower uncertainty interval of the quoted result for $\sigma(gg \to S)\;\times\;{\rm BR}(S \to \tau^+\tau^-)$. Also,  in the signal strength, $\Phi$ represents a fictitious SM Higgs boson with mass 95 GeV.

For completeness, we also report the signal strengths for the two other aforementioned anomalies, as follows:
%%%%%%%%%%%%%%%%%%%%%%%%%%%%%%
%A similar excess has also been reported in the di-photon final states by the CMS collaboration based on the full Run 1 and the first %Run 2 (35.9 fb$^{-1}$ ) data. The local and global significance are $2.8\sigma$
%and $1.3\sigma$ at $m_{\gamma\gamma} = 95.3$ GeV. This excess can be interpreted as a new resonance $h$, that decays to %two photons, i.e., $gg \to h \to \gamma \gamma$. The best fit value for the excess is~\cite{CMS:2018cyk} 
%\begin{eqnarray}\label{eq:excess_gg}
%\sigma(gg \to h){\rm BR}(h \to \gamma \gamma) = 0.058 \pm 0.019 ~\text{pb}
%\label{95 gamma gamma excess}
%\end{eqnarray}
%From the excess of events CMS obtained a signal strength of
\begin{eqnarray}\label{eq:excess_gg}
\mu^{\text{exp}}_{\gamma\gamma} = \frac{\sigma^{\text{exp}} (gg \to S \to \gamma\gamma)}{\sigma^{\text{SM}} (gg \to \Phi \to \gamma\gamma)} = 0.6 \pm 0.2
\end{eqnarray}
for $\gamma\gamma$ (see~\cite{CMS:2018cyk}) and 
%%%%%%%%%%%%%%%%%%%
%Moreover, another mild excess has been reported in the LEP experiment, that can be interpreted as $e^+ e^- \to Z S \to Z b %\bar{b}$. The signal corresponds to $2.3 \sigma$ local significance at $m_S = 98$ GeV. The excess observed at LEP can be %expressed in terms of a signal strength as~\cite{LEPWorkingGroupforHiggsbosonsearches:2003ing,ALEPH:2006tnd}
\begin{eqnarray}\label{eq:excess_bb}
\mu^{\text{exp}}_{bb} = \frac{\sigma^{\text{exp}} (e^+ e^- \to Z S \to Zb\bar{b})}{\sigma^{\text{SM}} (e^+ e^- \to Z \Phi \to Zb\bar{b})} = 0.117 \pm 0.057
\end{eqnarray}
for $b\bar b$ (see \cite{LEPWorkingGroupforHiggsbosonsearches:2003ing,ALEPH:2006tnd}). 

  In this work, in order to ascertain the CP state of the underlying spin-0 resonance (which, for illustrative purposes, we also assume to be a Higgs state),  we have adopted a model-agnostic explanation to all the excesses. As we particularly focus on the CP nature of this 95 GeV object, we have setup a field with Yukawa couplings to the third generation fermions, scaled by the corresponding SM Yukawa, while allowing in such an interaction for the simultaneous presence of both a $\gamma_5$ (i.e., a pseudoscalar interaction) and a unit matrix (i.e., a scalar interaction) structure, in both the $St\bar t$ and $S\tau^+\tau^-$ vertices, entering in production and decay, respectively. After exploring the parameter space of this simplified model by testing it against theoretical and experimental constraints, including fitting the described excesses, we look  at  unravelling the CP properties of the $S$ state at the LHC in the $\tau^+\tau^-$ channel using as diagnostic  observable  the signed acoplanarity  angle $\phi_{\rm CP}$ between the two $\tau$ lepton decay planes. In order to reconstruct the decay planes, we have used   both hadronic and semi-leptonic decays of the tau lepton, and used the Impact Parameter (IP) information~\cite{Berge:2014sra,Berge:2015nua} of the charged prong as well as
  the $\rho$-decay method~\cite{Bower:2002zx,Berge:2015nua,Hansen:2020ecn}.
%%%%%%%%%%%%%%%%%%%%%%%%%%%
%\emph{To reconstruct the decay planes we have used hadronic $\tau$ decays and impact %parameter(IP)~\cite{Berge:2014sra,Berge:2015nua} and $\rho^{\pm}$ decay plane %method~\cite{Bower:2002zx,Berge:2015nua,Han:2016bvf,Dutta:2021del}.} 
%%%%%%%%%%%%%%%%%%%%%%%%%%%
By then performing a  robust detector-level simulation, we show 
  that it is indeed possible to unambiguously differentiate a CP-even from a CP-odd as well as a  maximally CP-mixed signal at the LHC with 3000 fb$^{-1}$ of data per experiment, which is the expected final dataset size of the HL-LHC  \cite{Gianotti:2002xx,Apollinari:2015wtw}. 
  However, we deem this result  a conservative one since,
as we gather more data at the current LHC (Run 3),  the significance of the aforementioned excesses may well grow further, thereby stimulating especially dedicated analyses aimed at accessing the CP properties of this potential new resonance, which may well prove to offer an even better sensitivity than the one established  in this article. Yet, our findings are altogether already very encouraging in this respect. For the observed 125 GeV scalar, a similar CP characterization has been done by ATLAS~\cite{ATLAS:2022akr} and CMS collaboration~\cite{CMS:2021sdq}.
However, at the lower mass, the signal rates, backgrounds as well as kinematics are
different, and no dedicated CP study exists in this mass region. 
We believe this is important in order to test various hypotheses made in
the literature about the CP properties of possible interpretations.
  
Our paper is organised as follows. In the next section, we describe our  simplified model approach, including delineating its parameter space compatible with current constraints and explaining the $\gamma\gamma, \tau^+\tau^-$ and/or $b\bar b$ excesses. Then we move on to describe the construction of the mentioned CP-sensitive observable. Numerical results will follow, in turn preceding our conclusions.

\section{A Simplified Model to Address the Excesses}
We are interested in the study of CP properties of the observed spin-0 resonance, which might come from  various 
UV complete scenarios like the 2HDM, Next-to Minimal Supersymmetric Standard Model 
(NMSSM) and other scalar extensions of the SM. To remain model agnostic, we assumed a
 simplified model where, in addition to the SM spectrum, we consider only an additional scalar ($S$) 
 of mass 95 GeV. The scalar particle $S$ couples with the third generation fermions as follows: 
\begin{eqnarray}\label{eq:lag}
\mathcal{L}_{Sf\bar{f}} = -\rho^S_{f} \frac{m_f}{v}\Big(\cos \alpha \bar{f}f + i\sin \alpha \bar{f}\gamma_5 f\Big)S.
\end{eqnarray}    
We have introduced CP-Violating (CPV) Yukawa coupling with $\alpha$ being the CPV parameter. Here, $\alpha$ will 
be zero ($\pi/2$)  when the scalar $S$ is a CP-even (CP-odd) Higgs particle (e.g., in the 2HDM, this will be $H$ ($A$)). For simplicity, we have assumed 
a uniform CP mixing among generations. The dimensionless real parameters $\rho^S_f$ ($f=t,b,\tau$)
 are the Yukawa coupling modifiers. The observed SM-like CP-even Higgs boson of mass 125
  GeV can couple with $S$ depending on the Ultra-Violet (UV) completion. Our analysis will not depend on the coupling of
  $S$ with the SM Higgs boson. 
  
The coupling of the scalar $S$ with the weak neutral gauge boson, denoted as $g_{SVV}$, does not affect the LHC excess 
observed in the $\tau^+\tau^-$ and $\gamma\gamma$ channels. However, a non-zero $g_{SVV}$ is necessary to explain the LEP 
$b\bar b$ excess. Such coupling can appear if $S$ is part of a weak multiplet which mixes with the SM Higgs doublet. 
Since the observed 125 GeV Higgs boson at the LHC closely resembles the SM one, the mixing of the Higgs doublet 
cannot be too large, and we have considered $g_{SVV} < 0.3$.

\subsection{Production and Decay Modes}
Considering the large top mass and presuming $\rho^S_b$ not to be significantly large compared to $\rho^S_t$, we can obtain the gluon effective coupling described by a dimension five operator 
\begin{eqnarray}
\mathcal{L}_{hgg} = \rho_t^S \cos\alpha  ~A_S  ~S G^a_{\mu\nu} G^{a,\mu\nu} + \rho_t^S \sin\alpha ~A'_S ~S G^a_{\mu\nu} \widetilde{G}^{a,\mu\nu}
\end{eqnarray}
where $G_{\mu\nu}$ is the gluon field strength tensor and $\widetilde{G}_{\mu\nu}=\frac{1}{4}\epsilon_{\mu\nu\rho\sigma}G^{\rho\sigma}$ is the corresponding dual field strength tensor and  $A_S$ and $A'_S$ are effective couplings.
We incorporated Eq.~(\ref{eq:lag}) in 
\texttt{FeynRules}~\cite{Christensen:2008py,Alloul:2013bka} and used \texttt{SusHi}~\cite{Harlander:2012pb,Harlander:2016hcx} in order to estimate the Next-to-Next-to-Leading Order (NNLO) production
cross section via gluon fusion. The gluon fusion cross section in  the large top mass limit at NNLO is given by 
\begin{eqnarray}\label{eq:ggh}
\sigma(gg \to S) = {\rho_t^S}^2\Big( 76.35\cos^2\alpha + 176.32\sin^2\alpha \Big) ~\text{pb},
\end{eqnarray}
where the first term is the contribution from the CP-even part, and the second is the CP-odd contribution. 
The numerical coefficients are obtained from  \texttt{SusHi}~\cite{Harlander:2012pb,Harlander:2016hcx}.

%\subsection*{Decay Modes}
Finally, for our analysis, we have considered that the scalar  $S$  can decay via the following channels~\cite{Choi:2021nql}:
\begin{eqnarray}
S &\to& f \bar{f}, ~VV^* , ~gg,~\gamma\gamma  \textrm{ and }~Z\gamma . 
\end{eqnarray}

\subsection{Allowed Parameter Space}

\begin{figure}[tbh]
\begin{center}
  \includegraphics[width=0.49\linewidth]{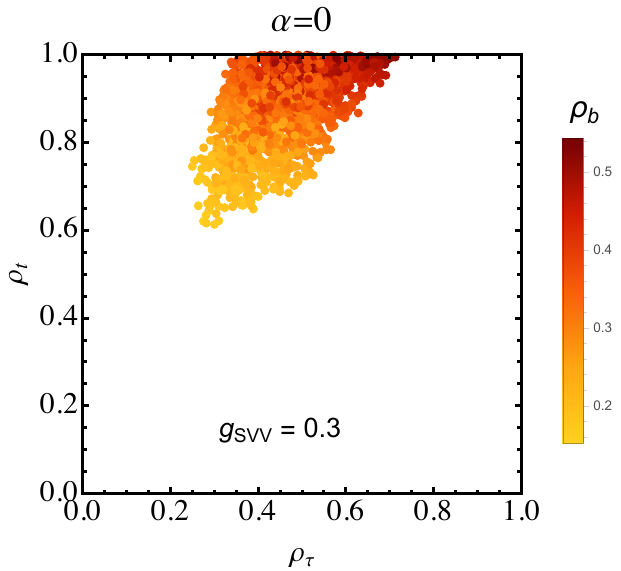}
  \includegraphics[width=0.49\linewidth]{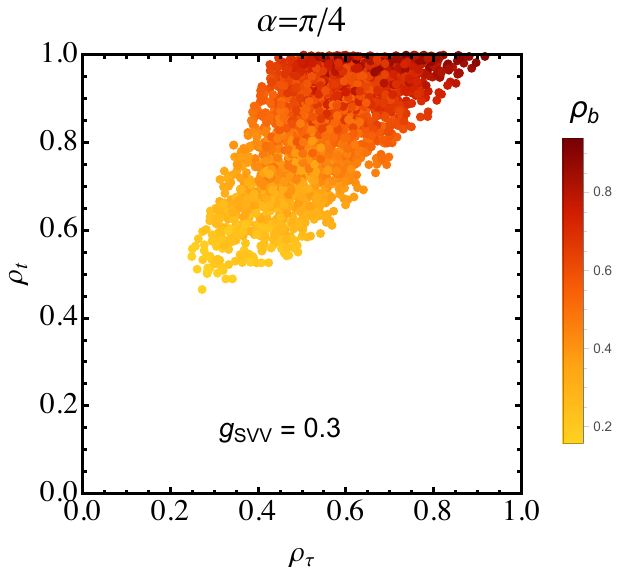}
    \includegraphics[width=0.49\linewidth]{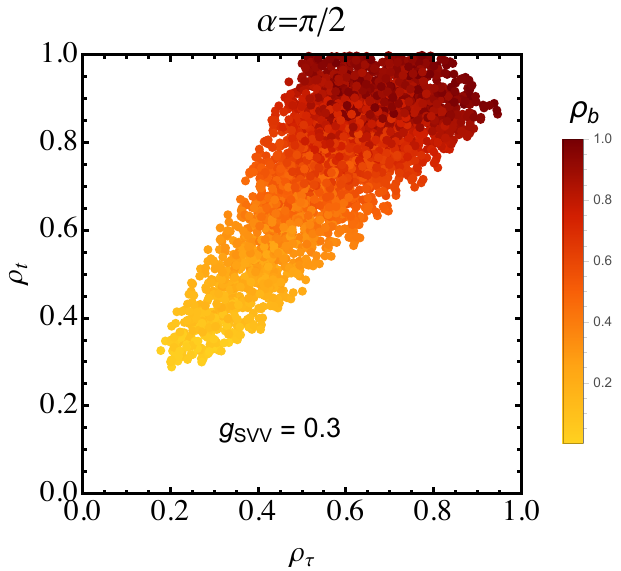}
  \caption{The allowed parameter space for the CP mixing angles $\alpha = 0, ~\pi/4$ and $\pi/2$. For the case of $\pi/2$, we cannot explain the LEP $b\bar{b}$ excess, and the allowed region is for the CMS $\tau^+\tau^-$ and $\gamma\gamma$ excesses. We can explain all the excesses for $\alpha = 0$ and $\pi/4$.}
  \label{fig:alpha}
\end{center}
\end{figure}

In order to explain all the anomalies, the scalar $S$ must be coupled to third-generation fermions as well as massive gauge bosons. We fix the CPV parameter $\alpha$ and scan the free parameters $\rho_{f}^{S}$ to 
satisfy the excesses defined in Eq.~(\ref{eq:excess_ta}), Eq.~(\ref{eq:excess_gg}) and Eq.~(\ref{eq:excess_bb}).
We have shown the parameter space that can simultaneously explain all the excesses in 
Fig.~\ref{fig:alpha}. 
For a CP-even scalar, the cross section is relatively small, as shown in Eq.~(\ref{eq:ggh}), and consequently, a large value
of $\rho_{t}$ is required, as depicted in the top-left panel of Fig.~\ref{fig:alpha}. For a CP-mixed state, the required values of 
$\rho_{t}$ decrease 
as the cross section increases due to a pseudoscalar component, the parameter is shown in the top-right panel of Fig.~\ref{fig:alpha}. In the bottom panel, we have considered a pure pseudoscalar model. Note that such a pure pseudoscalar 
scenario can not explain the LEP $b\bar b$ excess. Thus, the $g_{SVV}$ is redundant, and the viable parameter space does not 
depend on it. In all cases, the color represents the $\rho_{b}$ value. From the figure, it is evident that it is possible 
to satisfy all three excesses simultaneously. 

If we exclude the LEP excess, we can take $g_{SVV} = 0$. Such a choice alters the decay spectrum of $S$ as $S\to Z ff$ 
is not possible. In addition, it is possible to assume $\rho_{b}^{S} = 0 $ to explain the LHC excesses. Hence, to explain the 
LHC anomalies, as given in Eqs.~(\ref{eq:excess_ta})--(\ref{eq:excess_gg}), a lower value of $\rho_{t/\tau}^{S}$ compared to the parameter space shown in Fig~\ref{fig:alpha} will be 
enough, which corroborates \cite{Iguro:2022dok}. Note that, in Fig.~\ref{fig:alpha}, the parameter space 
does not depend on the sign of the angle $\alpha$, and thus, the results are identical for both $\pm\alpha$.

\section{Uncovering the CP Nature of the Resonance}
The CP transformation property of a scalar particle can be inferred from its decay to tau pairs since the correlation of the 
tau lepton polarization planes is governed by the CP properties of the parent particle. From the reconstruction of the decay 
planes of the tau leptons, it is then possible to infer the CP properties of the scalar particle. 

\subsection{Hadronic $\tau^\pm$ decays}
The differential decay width of a spin-0 boson into two fermions can be written as~\cite{Kramer:1993jn,Berge:2014sra,Berge:2014wta}
\be
d\Gamma_{S\to\tau^{+}\tau^{-}} \propto 1-\dfrac{\pi^{2}}{16}b(E_{+}) b(E_{-})\cos(\phi_{\rm CP}-2\alpha),
\ee 
where $b(E_{\pm})$ are the spectral function describing the spin analyzing power of a given decay mode~\cite{Tsai:1971vv,Berge:2014wta}. For $\tau^\pm \to \pi^\pm \nu$ decays, $b(E_{\pm}) = 1$ while, for other decay channels, 
it depends on the energy of the charged hadron at the respective tau-lepton rest frame. There are several methods 
~\cite{Bower:2002zx,Desch:2003rw,Berge:2013jra,Berge:2008wi,Berge:2008dr,Berge:2011ij,Berge:2014sra,Berge:2014wta,Berge:2015nua} to estimate $\phi_{\rm CP}$ and in this work we have used the Impact Parameter (IP)
 one~\cite{Berge:2014sra,Berge:2015nua} and the $\rho$ decay plane one~\cite{Bower:2002zx,Berge:2015nua,Han:2016bvf,Dutta:2021del}, for hadronic $\tau$ decays. For completeness, we will briefly describe the two methods next. 
  
\subsubsection{Impact Parameter Method}
 In the Impact Parameter (IP) method~\cite{Berge:2014sra,Berge:2015nua},  the $\tau$-lepton decay plane is constructed 
 using the 3-momentum vector ($\vec{q}^{\pm}$) and the 
 3-Dimensional (3D) impact parameter  vector ($\vec{n}_{\pm}$) of the charged particles. The method can be used for any 
 one-prong decay of the tau lepton and in this work we have implemented it for $S\to \tau^{+}\tau^{-} \to( \pi^{+} \nu_{\tau} )(\pi^{-}\bar\nu_{\tau})$ decay. 
 The IP vector in the laboratory 
 frame is defined as the distance of the closest approach of the charged particle’s ($\pi^{\pm}$) track to the 
 reconstructed Primary Vertex (PV)  of an event. It is possible to reconstruct the 3D IP vector using the distance between 
 the point of closest approach and 
 the primary vertex in the $x-y$-plane ($d_{0}$) and the $z$-coordinate of the point ($z_{0}$)~\cite{Hansen:2020ecn}. 
 We have used $d_{0}$ and $z_{0}$ and charged track momentum (highest $p_{T}$ charged track inside a $\tau$-tagged jet) to find out the IP vector using the procedure described in~\cite{Hansen:2020ecn}. 
 The normalized IP vector $\hat{n}$ is then converted to a four vector $n_{\pm}^{\mu} = (0,\hat{n}_{\pm})$ and boosted to Zero-Momentum Frame (ZMF), i.e., the $\pi^{+}\pi^{-}$ rest frame. The boosted IP four-vectors are denoted as $n_{\pm}^{*\mu}$ and we define the angle,
 \be
 \phi^{*} = \arccos(\hat n^{*}_{+_{\perp}}.~\hat n^{*}_{-_{\perp}}), 
 \ee
 where $\hat{n}^{*}_{\pm_{\perp}}$ is the normalized perpendicular component of $\vec{n}^{*}_{\pm}$ along the charged track $\pi^{\pm}$.
 To obtain a signed angle, we define the triple product, 
 $\mathcal{O} (= \hat q^{*}_{-}\cdot(\hat n^{*}_{+_{\perp}}.~\hat n^{*}_{-_{\perp}}))$ where $\hat q^{*}_{-}$ is the normalized 3-momentum vector of $\pi^{-}$ in the  ZMF. The CP observable is given by
 \be
\phi_{\rm CP} = \left\{
    \begin {aligned}
         & \phi^{*} \quad & \mathcal{O} \geq 0 \\
         & 2\pi-\phi^{*} \quad & \mathcal{O} < 0.
    \end{aligned}
\right.
\ee

 \subsubsection{Rho Decay Plane Method}
When the $\tau$ lepton decays via the rho meson, $\tau^\pm \to \rho^{\pm} \nu_{\tau} \to \pi^{\pm} \pi^{0} \nu_{\tau}$,
 we use the information from the charged and neutral pion to define a CP-odd observable.
In the  $\pi^{+}\pi^{-}$ rest frame, the acoplanarity angle $\phi^{*}$ is defined between the plane spanned by the 
$(\pi^{+},\pi^{0})$ and $(\pi^{-},\pi^{0})$ vectors. The acoplanarity angle is given by~\cite{Bower:2002zx,Berge:2015nua,Hansen:2020ecn},
\be
\phi^{*} = \arccos(\hat q^{~*0}_{+_{\perp}}.~\hat q^{~*0}_{-_{\perp}}), 
\ee
 where $\hat q^{~*0}_{\pm_{\perp}}$ is the normalized 3-momentum of the $\pi^{0}$ associated with $\pi^{\pm}$ in the $\pi^{+}\pi^{-}$
 rest-frame transverse to the direction of the associated charged pion. The signed angle $\tilde\phi^{*}$ is obtained using
the CP-odd triple correlation $\mathcal{O_{\rm CP}} (= \hat q^{*}_{-}\cdot(\hat q^{*0}_{+_{\perp}}\times~\hat q^{*0}_{-_{\perp}})$) and we define 
\be
\tilde\phi^{*} = \left\{
    \begin {aligned}
         & \phi^{*} \quad & \mathcal{O_{\rm CP}} \geq 0 \\
         & 2\pi-\phi^{*} \quad & \mathcal{O_{\rm CP}} < 0.
    \end{aligned}
\right.
\ee
In the $S\to \tau^{+}\tau^{-}\to (\pi^{+}\pi^{0}\nu_{\tau} )(\pi^{-}\pi^{0}\bar\nu_{\tau})$ decay, the spin analyzing power is 
proportional to $y_{\rho}=\dfrac{(E_{\pi^{+}}-E_{\pi^{0}})}{(E_{\pi^{+}}+E_{\pi^{0}})}\times\dfrac{(E_{\pi^{-}}-E_{\pi^{0}})}{(E_{\pi^{-}}+E_{\pi^{0}})}$ 
at the respective $\tau^{\pm}$ rest frame~\cite{Bower:2002zx}. Since $y_{\rho}$ can be both positive and negative and if we 
integrate over the charged and neutral pion moments, $\phi_{\rm CP}$ does not display CP sensitivity. We categorize the
events depending on $y_{\rho}$ and define the final CP-sensitive variable,
\be
\phi_{\rm CP}  =  \left\{
    \begin {aligned}
         &\tilde\phi^{*} +\pi \quad & y_{\rho} < 0 \\
         & \tilde\phi^{*}\quad & \textrm{ otherwise}.
    \end{aligned}
\right.
\ee
To estimate $y_{\rho}$, we have calculated the energies of the pions in the lab frame, as it is not possible to reconstruct the 
$\tau$ rest frames. The choice of lab frame energy does not significantly reduce the asymmetry in $\phi_{\rm CP}$ as long as the 
$\rho$ mesons are energetic ($p_{T}(\rho^{\pm}) > 20$ GeV)~\cite{Berge:2015nua} which is in general true for LHC.

\subsubsection{Combined IP and Rho Decay Plane Methods}
In the case of mixed decay, $\tau^+\tau^- \to (\pi\nu)( \rho\nu)$, we can combine the above-mentioned methods to 
define a CP-sensitive observable. This is possible if we consider the $\pi^{+}\pi^{-}$ rest frame for 
the $\rho$ Decay plane method. 

In the case of $\tau^{+}\tau^{-}\to( \rho^{+}\nu_{\tau})(\pi^{-}\bar\nu_{\tau})$ decays, the angle is defined by~\cite{Berge:2015nua},
\be
\phi^{*} = \arccos(\hat q^{~*0}_{+_{\perp}}.~\hat n^{~*}_{-_{\perp}}), 
\ee
where, $\hat q^{~*0}_{+}$ is the normalized 3-momentum of $\pi^0$ (originating from $\rho^+$) and $\hat n^{~*}_{-}$ is the IP vector of $\pi^{-}$ and  
all the vectors are in the $\pi^{+}\pi^{-}$ rest frame. As before, we can define a signed angle depending on the value of 
the triple product $\mathcal{O}_{3} (= \hat q^{~*}_{-}\cdot (\hat q^{~*0}_{+_{\perp}}\times~\hat n^{~*}_{-_{\perp}})$).
Instead, if $\tau^{+}$ decays to $\pi^{+} \nu_{\tau}$, we can define $\phi^{*} $
 and the triple product appropriately. Note that, in both cases of mixed decay, 
 we have considered the event classification based on $y^\rho_{\pm} \Big(= \dfrac{E_{\pi^{\pm}} - E_{\pi^{0}}}{E_{\pi^{\pm}} + E_{\pi^{0}}}\Big)$.
 
 \subsection{{Semi-leptonic Decays}} 
For the semi-leptonic channels, $S \to \tau^+ \tau^- \to (\ell^+ \nu_\ell \bar{\nu}\tau)(\tau_h^- \nu\tau)$ (and charge conjugate), where the leptons denote electrons and muons, the hadronic $\tau$ decay is reconstructed using either the IP method or the $\rho$ decay plane method, as described earlier. The leptonic side is treated as a one-prong decay, and we are using the impact parameter of the charged lepton for our analysis. 

Here, the CP-sensitive angle $\phi_{\rm CP}$ is defined in the $\ell$–$\pi^\pm$ ZMF between the hadronic decay plane normal and the lepton direction, with the sign determined from the corresponding triple product. Note that, to estimate $\phi_{\rm CP}$ in the case of a leptonic decay, we have applied an additional shift~\cite{Tsai:1971vv,Berge:2011ij} of $\pi$ due to the different sign in the spectral function.
While the spin analyzing power of the leptonic side is reduced by the two neutrinos in the decay, these modes benefit from clean lepton triggers and lower QCD backgrounds, making them a valuable complement to the fully hadronic decays\footnote{In principle, we can also use the fully leptonic decay, but the corresponding distribution turns out to be the worst among all the possible modes and thus we have not used this mode in our analysis.}.

\subsection{Event Simulation and Selection}

 We implemented our simplified model for event simulation in \texttt{FeynRules-2.3}~\cite{Alloul:2013bka}. 
The signal and the background events are generated using 
\texttt{MadGraph5-aMC@NLO-3.5.3} \cite{Alwall:2011uj,Alwall:2014hca}. To handle the $\tau$ polarization, 
we have used the \texttt{TauDecay}~\cite{Hagiwara:2012vz} package in \texttt{MadGraph5-aMC@NLO-3.5.3}. 
For showering and hadronization we used \texttt{PYTHIA-8.3} \cite{Sjostrand:2014zea, Bierlich:2022pfr} while detector 
effects were simulated using  \texttt{Delphes-3.5.0} \cite{deFavereau:2013fsa}. We use the anti-kt algorithm \cite{Cacciari:2008gp} 
for jet clustering with radius parameter $R = 0.5$ and $p_T(j) > 20$ GeV.  For $\tau$-tagging, we have used a medium 
 tag point with $70\%$ tagging and $5\times 10^{-3}$ mis-tagging efficiency. 
 
 % We follow Ref.~\cite{CMS:2022goy} for event selection. 
 % The signal events are classified as two $\tau$-tagged jets with $p_{T} > 40$ GeV and $|\eta|<2.1$. We veto events with any $b$-tagged jets, and we have considered events having $p_T^{\tau^+\tau^-} > 200$ GeV as shown in Fig.~9 in Ref.~\cite{CMS:2022goy}.
 %
% For semi-leptonic events, electrons (muons) should have $|\eta| < 2.1$, and need to satisfy isolation criterion of $I_{\text{rel}} < 0.15$. For events selected by a single-$e$ trigger, the $p_T$ threshold on the electron is raised to 26, 28, or 33~GeV for the data-taking years 2016, 2017, and 2018, respectively. For a single-muon trigger, the $p_T$ requirement is increased to 23~GeV in 2016 and to 25~GeV in 2017--2018. The hadronically decaying tau ($\tau_h$) candidate is required to have $|\eta| < 2.3$. A $p_T$ threshold of 35~GeV (32~GeV) is applied when the event is selected by an electron--$\tau_h$ ($\mu$--$\tau_h$) pair trigger, while a threshold of 30~GeV is applied when the event is selected by a single-electron or single-muon trigger.
% {\textcolor{red}{SM: What's this listing doing here?!}}
%
We follow Ref.~\cite{CMS:2022goy} for event selection. For the hadronic channel, the signal events are classified as two $\tau$-tagged jets with $p_{T} > 40$ GeV and $|\eta|<2.1$. 
 For the semi-leptonic events, electrons (muons) should have $|\eta| < 2.1$, and need to satisfy isolation criterion of $I_{\text{rel}} < 0.15$. The electrons  (muons) with $p_T >$ 25 GeV (20) GeV is required for the $e-\tau_h$ ($\mu-\tau_h$) mode. The hadronically decaying tau ($\tau_h$) candidate is required to have $|\eta| < 2.3$. A $p_T$ threshold of 35~GeV (32~GeV) is applied when the event is selected by an $e-\tau_h$ ($\mu-\tau_h$) pair modes.
 For both the hadronic and the semi-leptonoc modes, We veto events with any $b$-tagged jets, and we have considered events having $p_T^{\tau^+\tau^-} > 200$ GeV as shown in the right panel of Fig.~9 in Ref.~\cite{CMS:2022goy}, where $p_T^{\tau^+\tau^-}$ is obtained by the vectorial $p_T$ sum of the visible decay products of tau leptons. Such  $p_T^{\tau^+\tau^-}$ criteria help to substantially minimize the background events.

  We obtain the charged and neutral pions from the $\tau$-jet constituents. From all the tracks within each $\tau$-jet, the track with the highest $p_T$ is considered to be the charged pion. Neutral pions decay promptly into a pair of photons and therefore deposit their energies in the electromagnetic calorimeter. From all the energy deposits, called towers, we estimate the one with the highest $E_T$ and assign the tower to the neutral pion. For the IP-$\rho$ combination method, the charged pion track with higher $p_T$ is associated with the two-body decay of the $\tau^\pm$ ($\tau^\pm \to \pi^\pm \nu_\tau$). In the case of the semi-leptonic mode, the track associated with the lepton is considered while, for the hadronic part, we have used either the IP or the $\rho$-decay method.

 The background for such signal at the LHC is dominated by  $Z\to \tau^+\tau^-$, $W$ + jets, $t\bar t$ and QCD multi-jet events. 
We have generated sample background data to understand the nature of the corresponding distribution for our CP-sensitive observable. 
As we will discuss later, we have used actual data provided by the CMS collaboration through \texttt{HEPDATA}~\cite{Maguire:2017ypu,hepdata.128147} and rescaled the background accordingly for a different luminosity.

 \section{Results}
 \begin{figure}[h!]
 \includegraphics[width=8cm,angle=0]{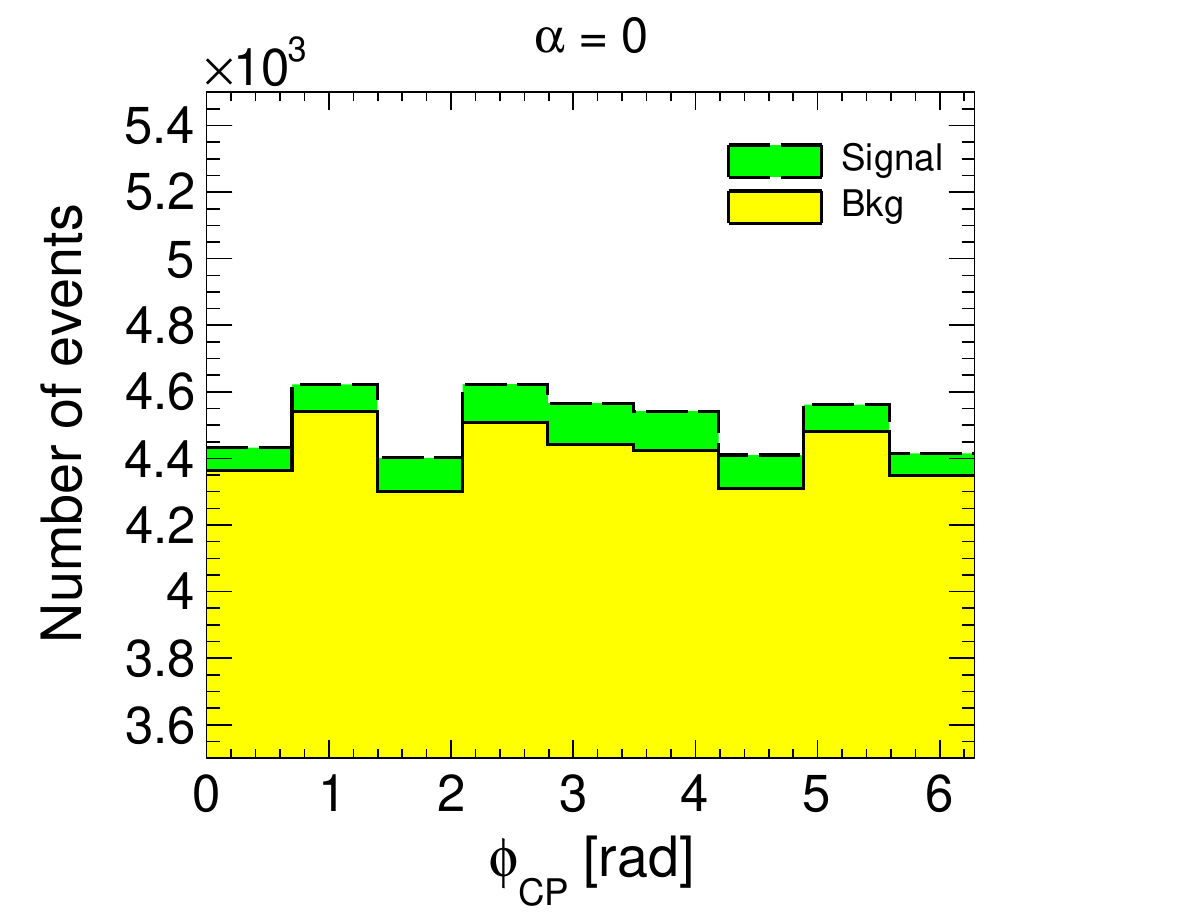}
  \includegraphics[width=8cm,angle=0]{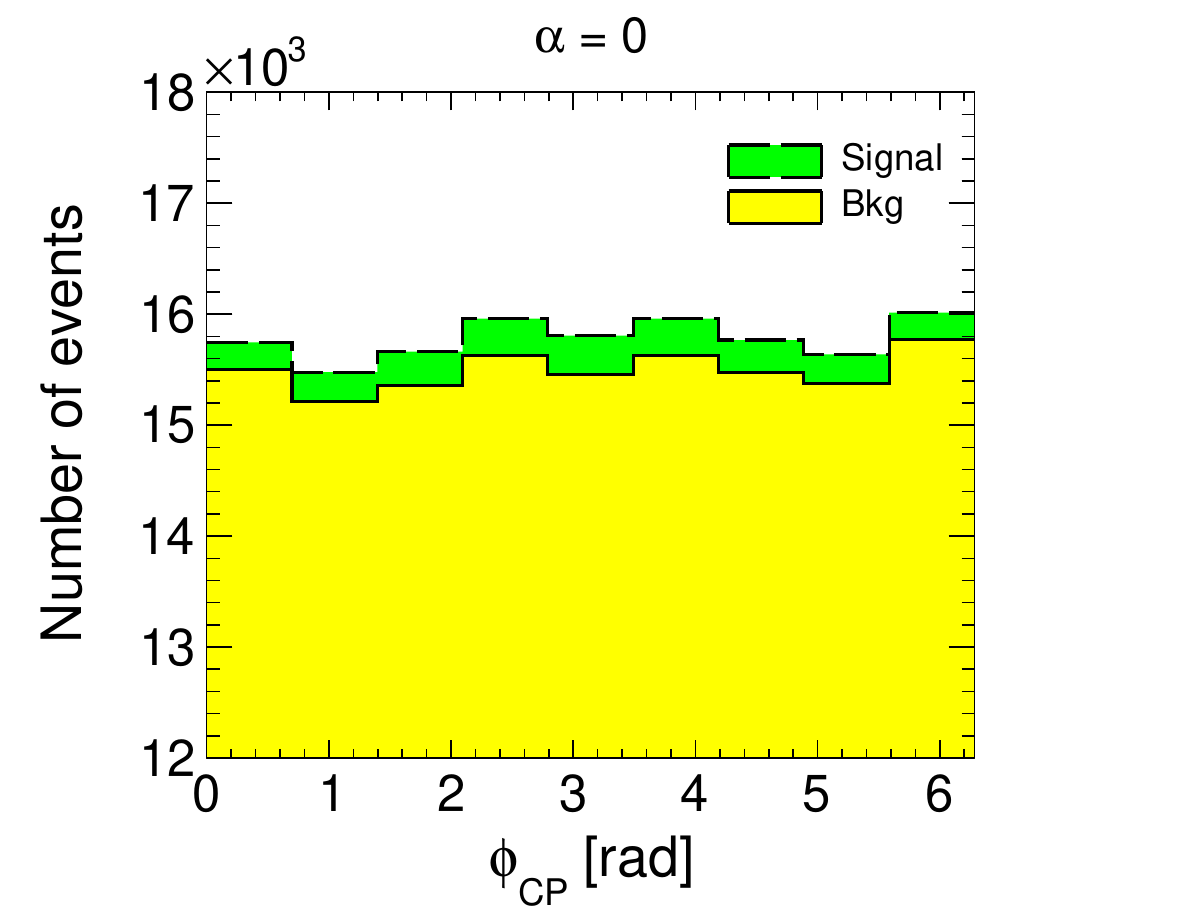}
 \includegraphics[width=8cm,angle=0]{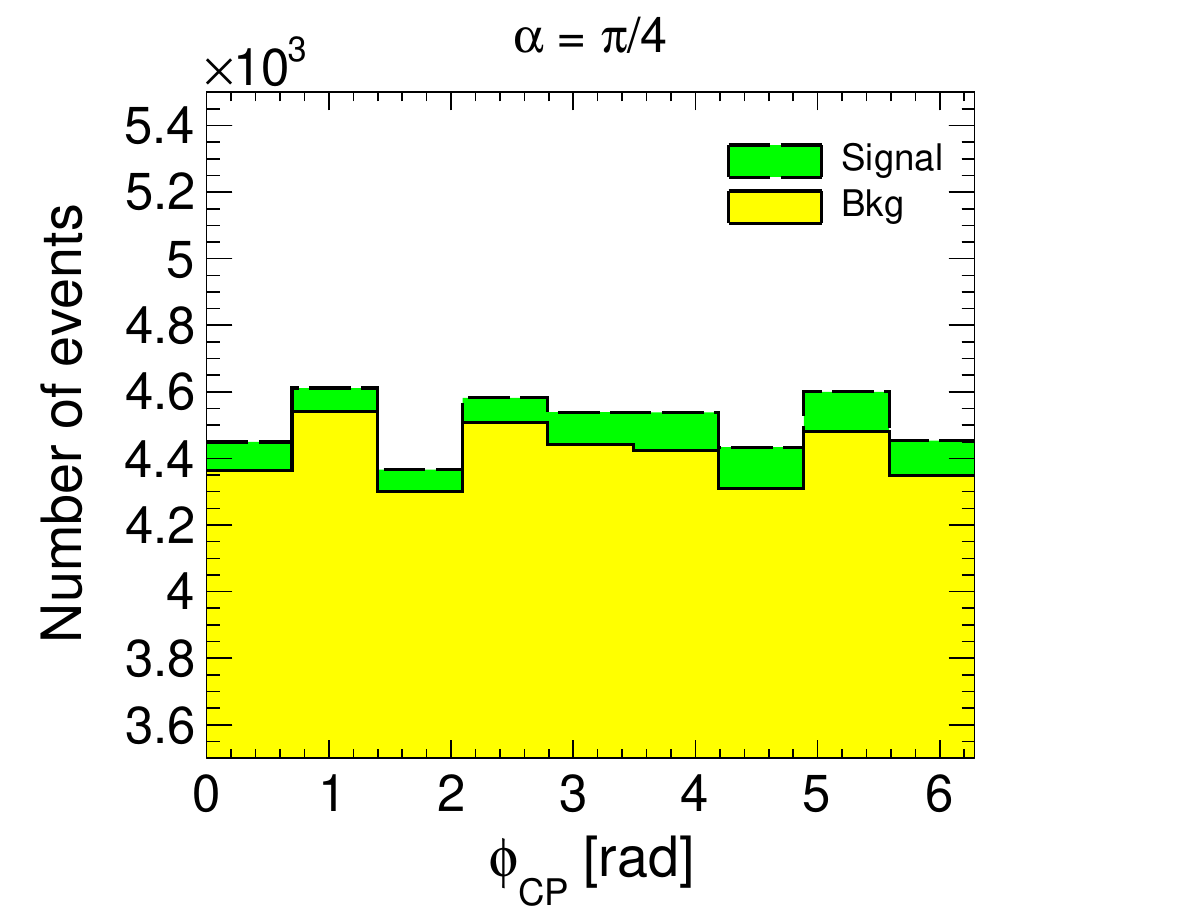}
 \includegraphics[width=8cm,angle=0]{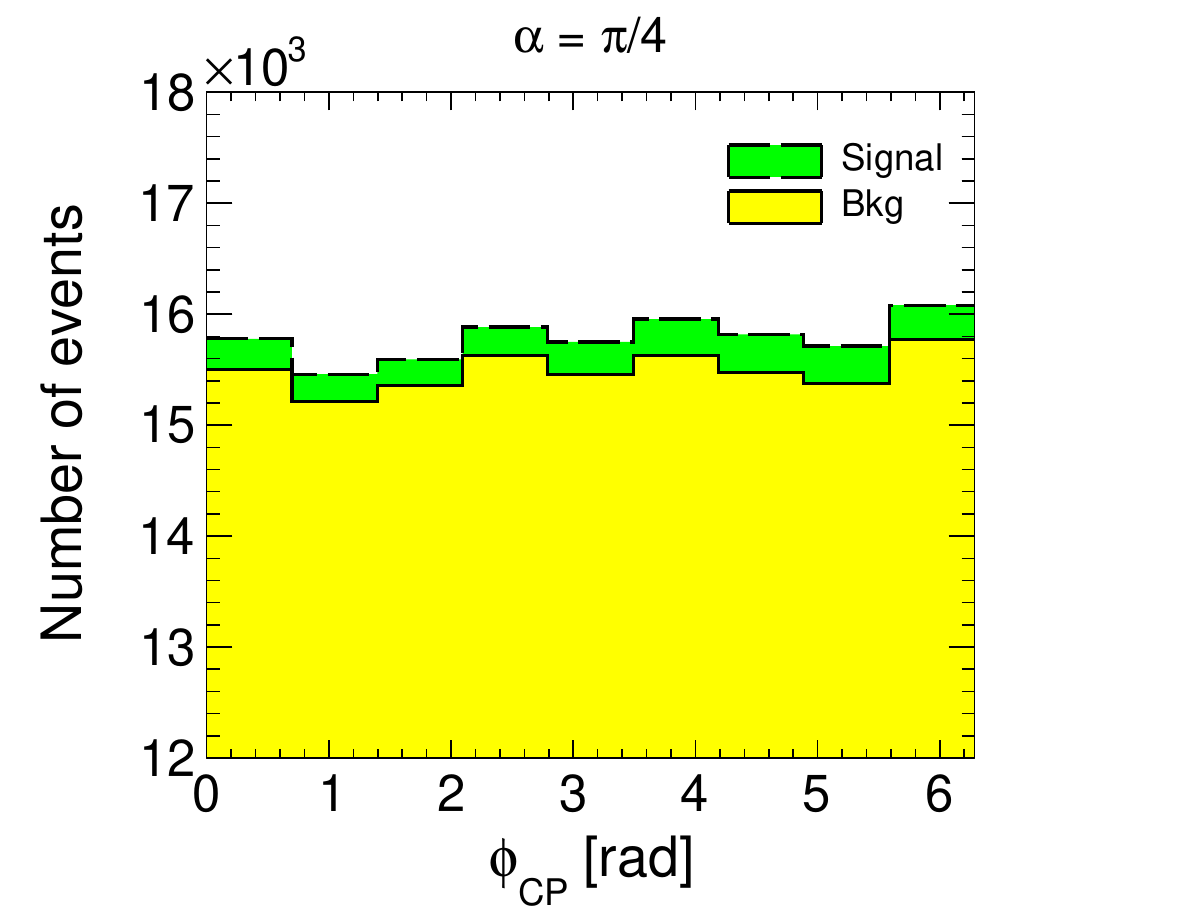}
 \includegraphics[width=8cm,angle=0]{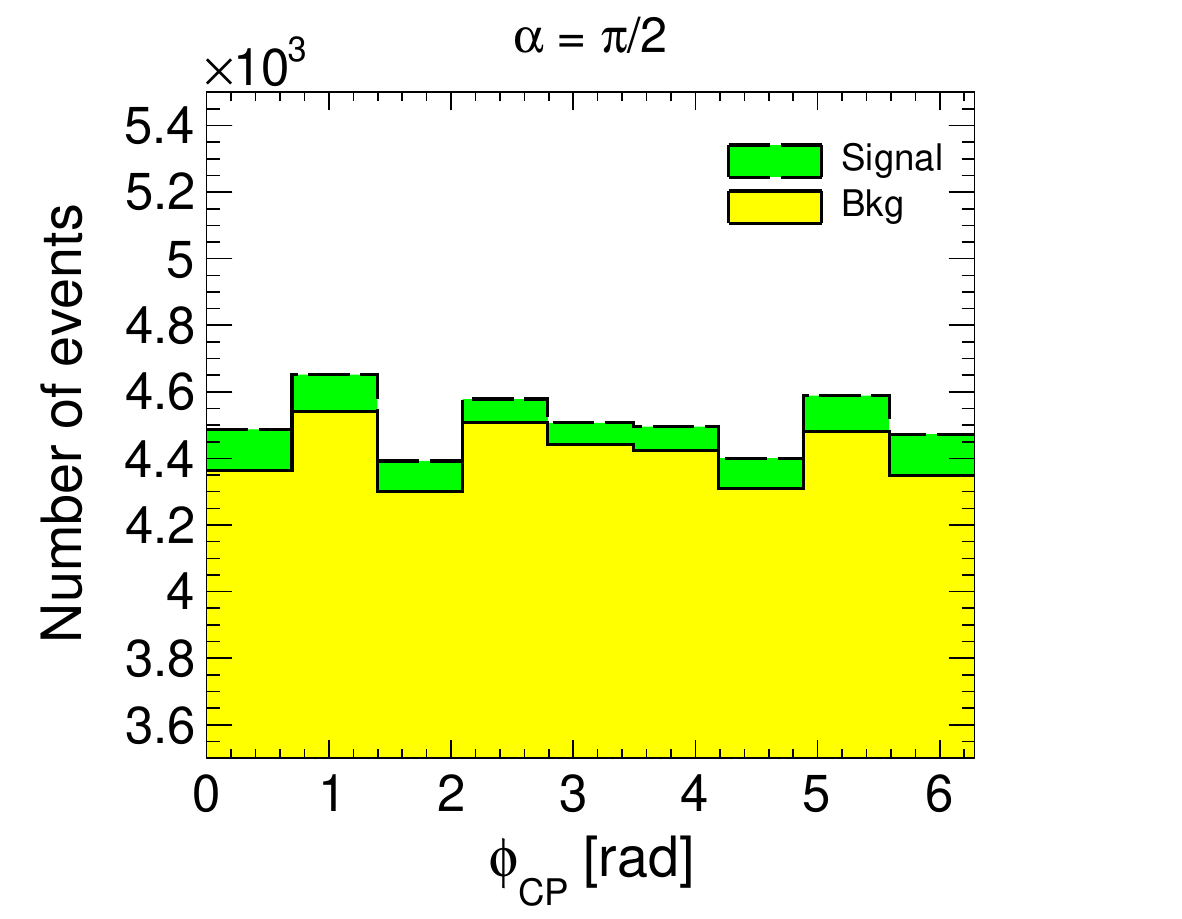}
 \includegraphics[width=8cm,angle=0]{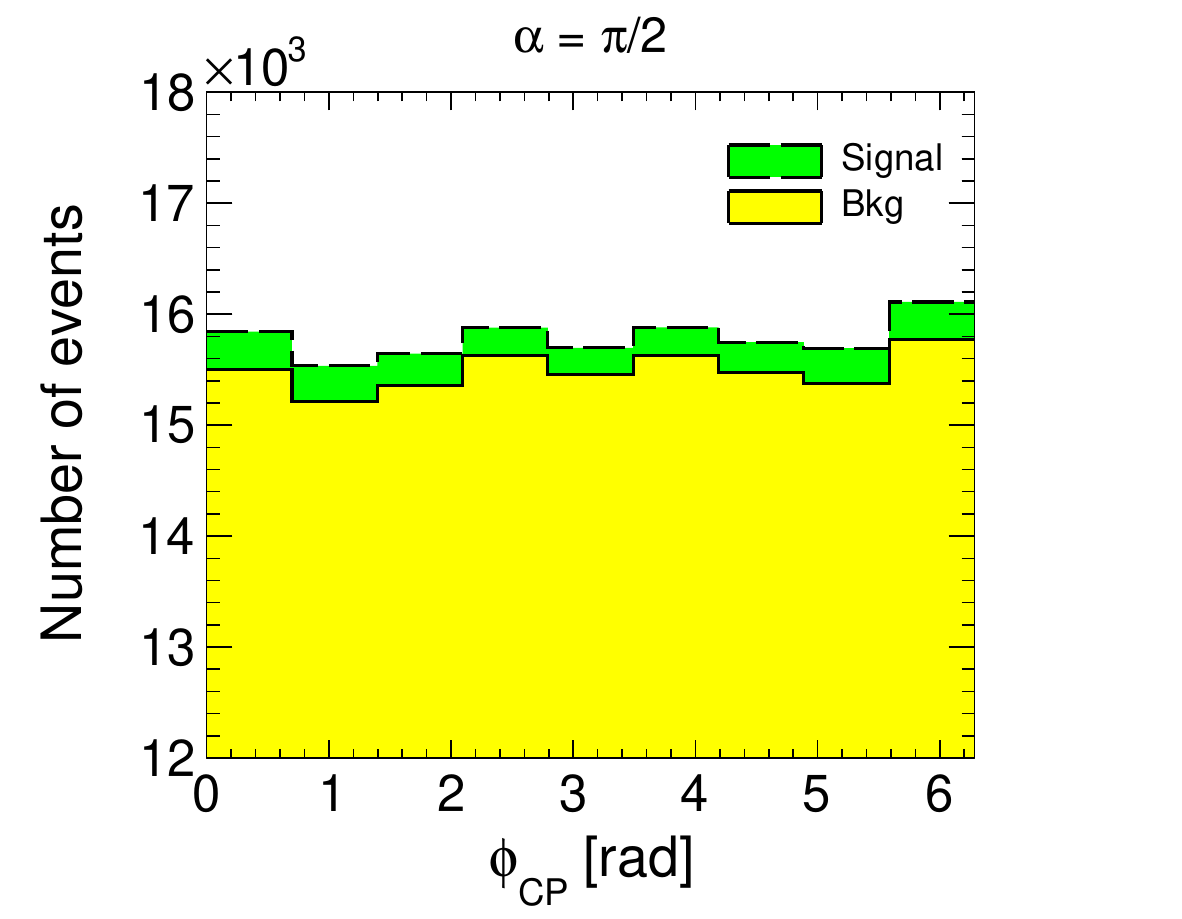}
 \caption{Signal and background distribution of our CP-sensitive observable for combining all the hadronic (left panels) and semi-leptonic (right panels) modes with an integrated luminosity of 3000 fb$^{-1}$. The top panel shows the distribution for a CP-even 95 GeV scalar, while the middle panel shows the same for a maximally CP-mixed state, and the CP-odd distribution is shown in the bottom panel.}
 \label{fig:phicp-139}
 \end{figure}
%
%
% The distribution peaks at a particular value of $\phi_{\rm CP}$ for a given CP admixture, whereas the background remains primarily flat. This enables estimating the  CP nature from the $\phi_{\rm CP}$  distribution. 

\begin{figure}[t!]
 \begin{center}
 \includegraphics[width=0.49\linewidth]{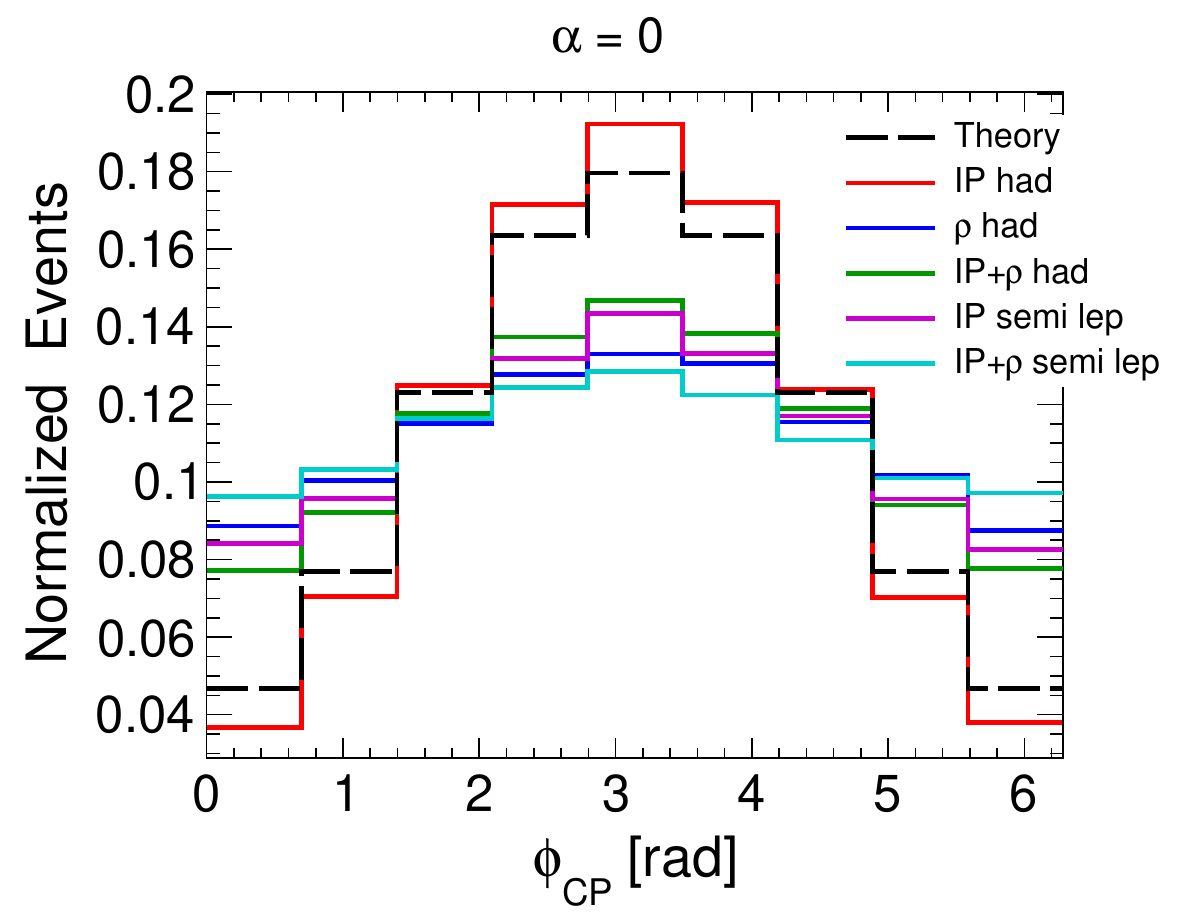}
\includegraphics[width=0.49\linewidth]{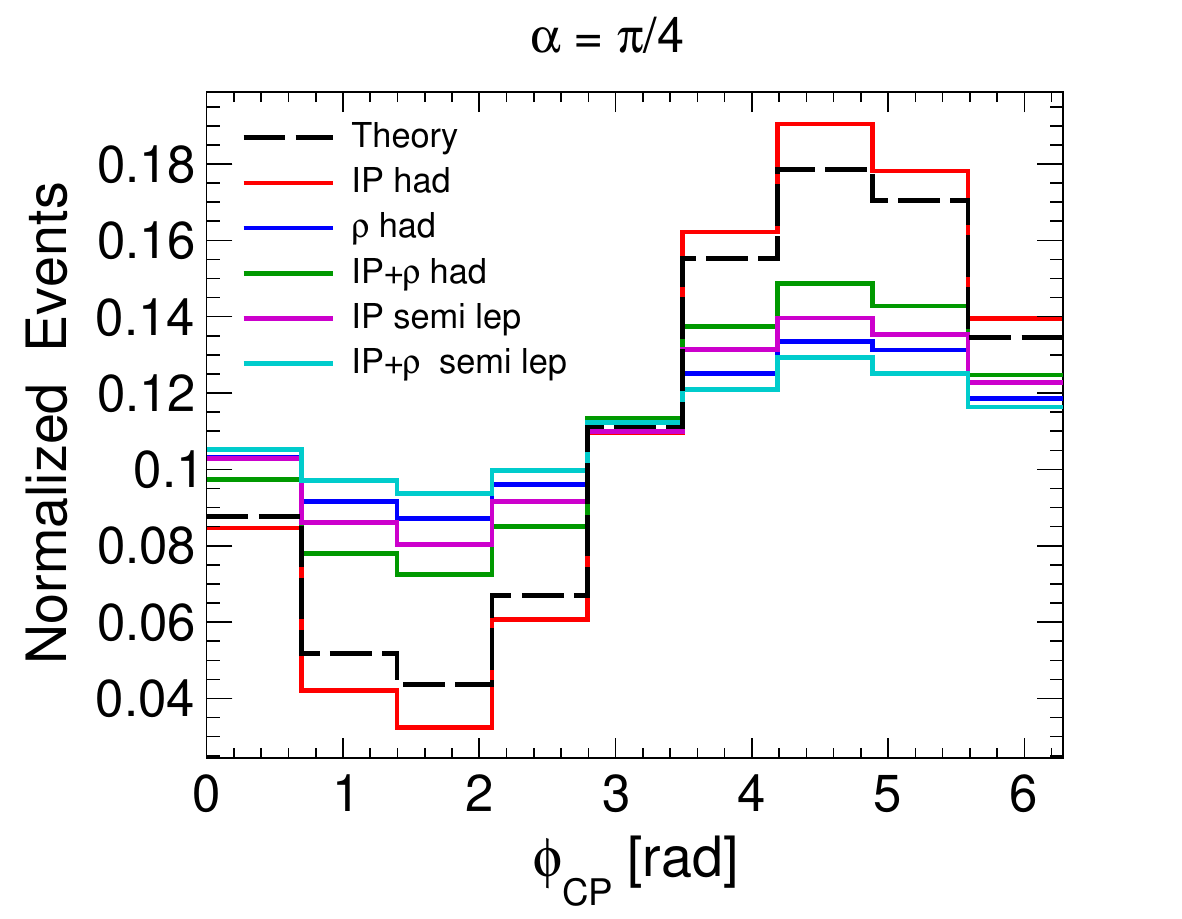}
\includegraphics[width=0.49\linewidth]{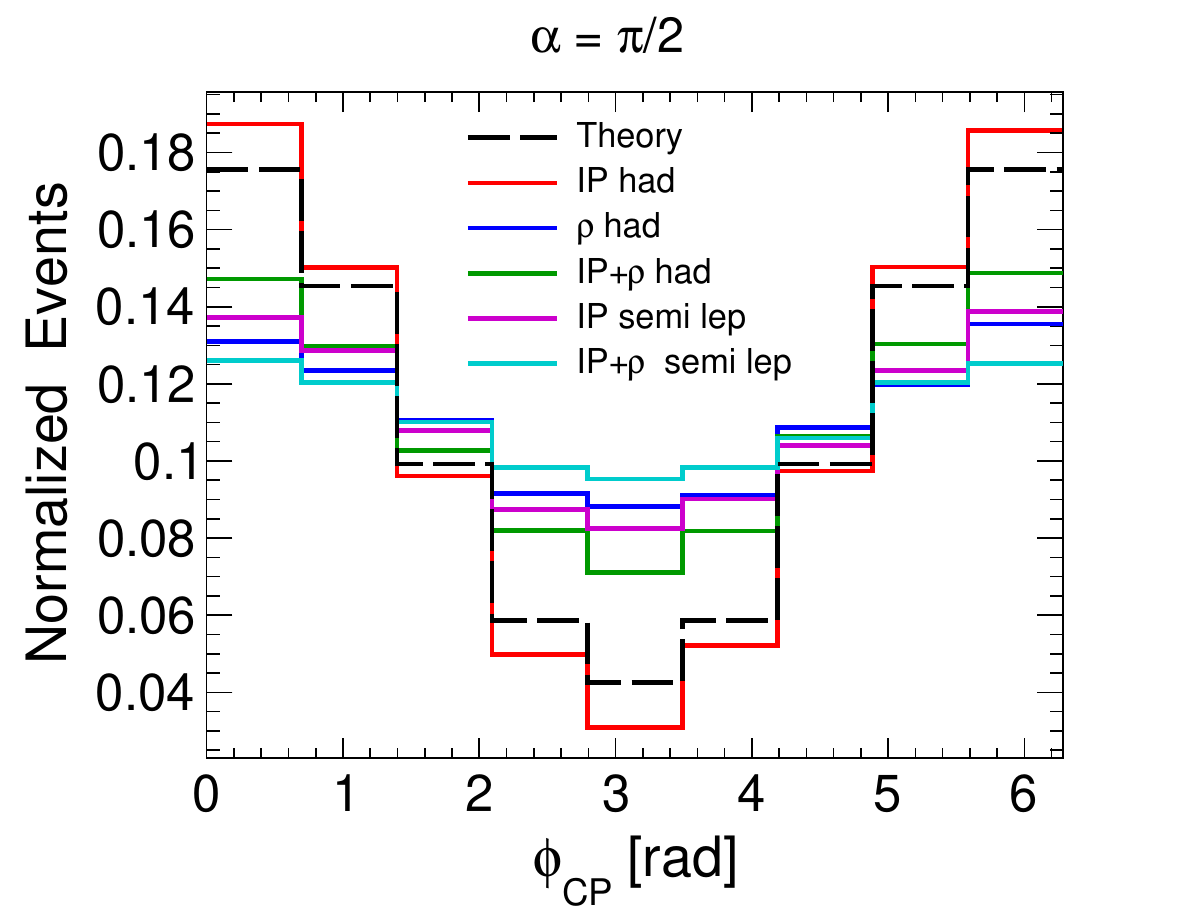}
 \caption{Signal and theoretical distribution fo the CP observable $\phi_{\rm CP}$ scaled to HL-LHC from the observed data at 139 fb$^{-1}$. The top panels show the distribution for CP-even (top-left) and maximally CP-mixed (top-right) scalar, and the bottom panels show the distribution for a CP-odd state. The hadronic IP method provides the best reconstruction and closely resembles the theoretical distribution, whereas the other methods show the exact nature of the distribution with subdued extrema. If $\alpha$ becomes $-\pi/4$, then the peak of the distribution will be at $\phi_{\rm CP} = \pi/2$.}
 \label{fig:norm-signal}
 \end{center}
 \end{figure}
 
% Now, we present the $\phi_{\rm CP}$ distribution for different CP states of the 95 GeV scalar and the background events. 
Now, we show how to access the CP properties of the spin-0  with particle with mass around 95 GeV at the LHC.
 For a realistic representation of signal and background events, we have used the present CMS data of the 
 $\tau^+\tau^-$ excess~\cite{CMS:2018cyk,CMS:2023yay}. 
 The data is taken from \texttt{HEPDATA} % \cite{Maguire:2017ypu,hepdata.128147}. 
 and for our analysis, we have selected the events 
 for which the invariant mass $m_{\tau^+\tau^-}$ lies within $60-120$ GeV  
 with $p_T^{\tau^+\tau^-} > 200$ GeV.
To demonstrate the CP nature of the spin-0 object, we obtain the $\phi_{\rm CP} $ distribution for both signal and background events. 
% The distribution peaks at a particular value of $\phi_{\rm CP}$ for a given CP admixture, whereas the background remains primarily flat. This enables estimating the  CP nature from the $\phi_{\rm CP}$  distribution. 
We found that, with an integrated luminosity of 138 fb$^{-1}$, the total event distribution remains within statistical uncertainty of the background ($\sqrt{N_{B}}$) and that the situation is not considerably better at the end of Run 3. Hence, we considered the  HL-LHC~\cite{Gianotti:2002xx,Apollinari:2015wtw} and appropriately rescaled the existing data to an integrated luminosity of 3000 fb$^{-1}$. 
 The $\phi_{\rm CP} $ distribution for signal and background events for the HL-LHC is shown in Fig.~\ref{fig:phicp-139},  where background events are shown in yellow, whereas the signal events are shown in green. The combined hadronic events are shown in the left panel, whereas the right panel depicts semi-leptonic events after combining all the modes. The top, middle and bottom panels in Fig.~\ref{fig:phicp-139} depicts the distribution for a CP-even, maximally CP-mixed and CP-odd state, respectively. As anticipated, the background events exhibit a near-uniform distribution in all cases. As theoretically expected, the signal distribution peaks at $\phi_{\rm CP} = \pi (0\textrm{ and }2\pi)$ for a CP-even(odd) spin-0 state, whereas for the CP-mixed one, the peak is at $3\pi/2$. From all the figures, it is evident that, for a given CP nature of the resonance, the cumulative number of signal events rises beyond the background uncertainty, thus signalling the CP properties of the new particle.

\begin{figure}[t]
\begin{center}
 \includegraphics[width=12cm,angle=0]{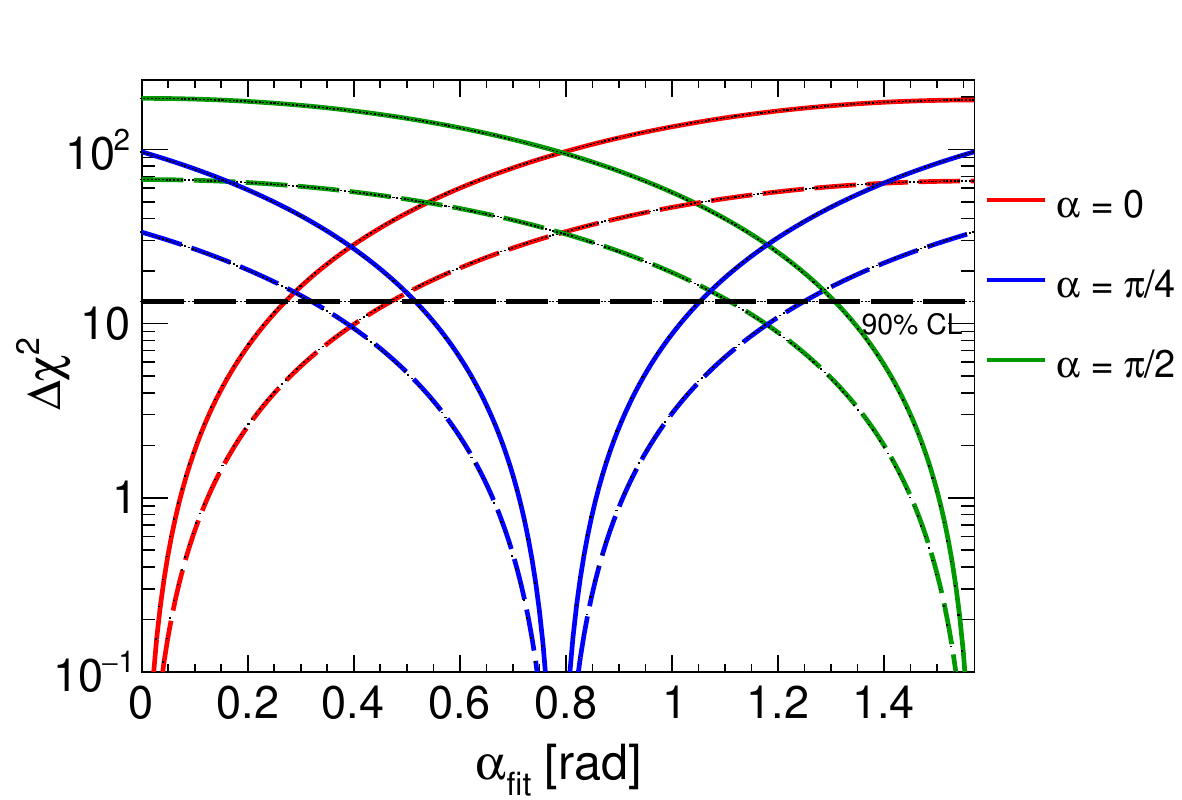}
 \caption{Plot of $\Delta\chi^2$ as a function of fitting angle $\alpha_{fit}$ assuming that the 95 GeV scalar to be CP-even(red), maximal CP-mixed(blue) and CP-odd(green). The solid(dashed) line corresponds to background systematics assumed to be 0.5\%(1\%). The black horizontal line at $\Delta \chi^2 = 13.36$ shows the 90\% Confidence Level (CL) interval. }
 \label{fig:chisq}
\end{center}
 \end{figure}

To quantify our results, we use the $\chi^2$ defined as
\begin{equation}\label{eq:chisq}
    \chi^2(\alpha) = \sum_{\rm Modes}\quad  \sum_{i\in \rm{Bins}} \dfrac{\left(S_i^{\alpha_H} - \frac{n_S}{\Gamma}\frac{d\Gamma}{d\phi_{\rm CP}}(\alpha)\right)^2}{(\delta S_i)^2 + (\delta_{sys}^2)},
\end{equation}
where $\alpha_H$ is the assumed hypothesis (i.e., the CP nature) for the 95 GeV state and $S_i^{\alpha_H}$ denotes the signal events in the $i^{th}$ bin. The input parameter $\alpha$ provides the theoretical estimation of the CP observable $\phi_{CP}$  normalized to the total number of signal events, $n_S$. For each bin, the variance includes the statistical component defined as $\delta S_i = \sqrt{S_i}$ and an uncorrelated systematic uncertainty corresponding to 0.5\% or 1\% of the expected background in that bin. Finally, the summation over modes includes all the semi-leptonic and hadronic $\tau^+\tau^-$ divided into IP, $\rho$ and combined IP-$\rho$ modes. The normalized binned distribution of signal events $(S_i^{\alpha_H})$ as well as the theoretical distribution  $\left(\dfrac{n_S}{\Gamma}\dfrac{d\Gamma}{d\phi_{\rm CP}}(\alpha)\right)$ for various channels is shown in Fig.~\ref{fig:norm-signal} for a CP-even (top-left), maximally mixed-CP state (top-right) as well as CP-odd (bottom). As expected, the hadronic IP method yields the most accurate distribution, closely matching the theoretical distribution, while the other methods capture the overall shape, but with maxima and minima systematically suppressed relative to the theoretical distribution.

In Fig.~\ref{fig:chisq}, we present our main result: the distribution of $\Delta\chi^2$ as a function of $\alpha_{\text{fit}}$. The red, blue and green lines correspond to the assumed hypotheses that the 95 GeV scalar is CP-even, maximally CP-mixed, and CP-odd, respectively. The solid (dashed) curves represent $\delta_{\text{sys}} = 0.5\% \ (1\%)$ of the total background, which enters the $\chi^2$ calculation as defined in Eq.~(\ref{eq:chisq}). The signal distribution is binned with $N_{\text{bin}} = 9$ and the $90\%$ CL interval corresponds to $\Delta\chi^2 = 13.36$, shown as a black horizontal line in Fig.~\ref{fig:chisq}. We find that, at the HL-LHC, it will be possible to ascertain the CP nature of the 95 GeV scalar within $\pm 0.27 \ (0.47)$ radians of the true hypothesis at $90\%$ CL, depending on the assumed background systematics.

\section{Conclusions}

In summary, under the assumption that several anomalies seen in the invariant mass of $\gamma\gamma$, $\tau^+\tau^-$ (from LHC data) and $b\bar b$ (from LEP data) events are due to a single spin-0 object with mass approximately 95 GeV having Yukawa couplings to third generation fermions, we have performed a detector level analysis aimed at extracting the CP properties of such an object. We have done so limitedly to the $\tau^+\tau^-$ final state, as it offers the possibility of accessing the corresponding CP quantum numbers, whichever the  $\tau$ decay, whether leptonic or hadronic, through the relative angle between the two decay planes defined through the visible particles emerging from the $\tau$ lepton decays. Our approach has been based on a simplified model, which is characterized by the relevant Yukawa interactions being defined in terms of correcting factors to the corresponding SM vertex strengths while allowing for both a scalar and pseudoscalar interaction as well as any superposition thereof. 
 
We have found that, under the assumption that the $\gamma\gamma$ and $\tau^+\tau^-$ anomalies will persist throughout the LHC era, the HL-LHC, in standard energy and luminosity configurations, will offer the chance to extract the CP properties of such a new spin-0 resonance. Once the latter is combined with cross section and BR information, it would then be possible to map the results obtained in our model-independent parameterization into well-defined models like, e.g., a 2HDM of whichever Type, so long that the latter can host a suitable 95 GeV state in its Higgs particle spectrum.

We therefore advocate experimental analyses by ATLAS and CMS aimed at implementing our approach, or indeed improved versions of it, with upcoming datasets at the CERN machine. 

 \section*{Acknowledgements}
 T.M. and P.S. would like to thank Artur Gottmann and Ravindra Kumar Verma for some useful discussions. 
 T.M. is supported by the BITS Pilani Grant NFSG/PIL/2023/P3801. S.M. is supported in part through the NExT Institute and  STFC Consolidated Grant No. ST/L000296/1. P.S. is supported by Basic Science Research Program through the National Research Foundation of Korea (NRF) funded by the Ministry of Education through the Center for Quantum Spacetime (CQUeST) of Sogang University (RS-2020-NR049598). P.S. would also like to thank the
 KIAS Center for Advanced Computation for providing computing resources.

%***********************************
%***********************************
\bibliographystyle{JHEP}
\bibliography{reference}

\providecommand{\href}[2]{#2}\begingroup\raggedright\begin{thebibliography}{10}

\bibitem{ATLAS:2012yve}
{\scshape ATLAS} collaboration, \emph{{Observation of a new particle in the
  search for the Standard Model Higgs boson with the ATLAS detector at the
  LHC}}, \href{https://doi.org/10.1016/j.physletb.2012.08.020}{\emph{Phys.
  Lett. B} {\bfseries 716} (2012) 1}
  [\href{https://arxiv.org/abs/1207.7214}{{\ttfamily 1207.7214}}].

\bibitem{CMS:2012qbp}
{\scshape CMS} collaboration, \emph{{Observation of a New Boson at a Mass of
  125 GeV with the CMS Experiment at the LHC}},
  \href{https://doi.org/10.1016/j.physletb.2012.08.021}{\emph{Phys. Lett. B}
  {\bfseries 716} (2012) 30} [\href{https://arxiv.org/abs/1207.7235}{{\ttfamily
  1207.7235}}].

\bibitem{Gunion:1992hs}
J.~F. Gunion, H.~E. Haber, G.~L. Kane and S.~Dawson, \emph{{Errata for the
  Higgs hunter's guide}},
  \href{https://arxiv.org/abs/hep-ph/9302272}{{\ttfamily hep-ph/9302272}}.

\bibitem{Branco:2011iw}
G.~C. Branco, P.~M. Ferreira, L.~Lavoura, M.~N. Rebelo, M.~Sher and J.~P.
  Silva, \emph{{Theory and phenomenology of two-Higgs-doublet models}},
  \href{https://doi.org/10.1016/j.physrep.2012.02.002}{\emph{Phys. Rept.}
  {\bfseries 516} (2012) 1} [\href{https://arxiv.org/abs/1106.0034}{{\ttfamily
  1106.0034}}].

\bibitem{Moretti:2019ulc}
S.~Moretti and S.~Khalil, \emph{{Supersymmetry Beyond Minimality: From Theory
  to Experiment}}. CRC Press, 2019.

\bibitem{DeCurtis:2018zvh}
S.~De~Curtis, L.~Delle~Rose, S.~Moretti and K.~Yagyu, \emph{{A Concrete
  Composite 2-Higgs Doublet Model}},
  \href{https://doi.org/10.1007/JHEP12(2018)051}{\emph{JHEP} {\bfseries 12}
  (2018) 051} [\href{https://arxiv.org/abs/1810.06465}{{\ttfamily
  1810.06465}}].

\bibitem{DeCurtis:2018iqd}
S.~De~Curtis, L.~Delle~Rose, S.~Moretti and K.~Yagyu, \emph{{Supersymmetry
  versus Compositeness: 2HDMs tell the story}},
  \href{https://doi.org/10.1016/j.physletb.2018.09.042}{\emph{Phys. Lett. B}
  {\bfseries 786} (2018) 189}
  [\href{https://arxiv.org/abs/1803.01865}{{\ttfamily 1803.01865}}].

\bibitem{CMS:2018cyk}
{\scshape CMS} collaboration, \emph{{Search for a standard model-like Higgs
  boson in the mass range between 70 and 110 GeV in the diphoton final state in
  proton-proton collisions at $\sqrt{s}=$ 8 and 13 TeV}},
  \href{https://doi.org/10.1016/j.physletb.2019.03.064}{\emph{Phys. Lett. B}
  {\bfseries 793} (2019) 320}
  [\href{https://arxiv.org/abs/1811.08459}{{\ttfamily 1811.08459}}].

\bibitem{CMS:2022goy}
{\scshape CMS} collaboration, \emph{{Searches for additional Higgs bosons and
  for vector leptoquarks in $\tau\tau$ final states in proton-proton collisions
  at $\sqrt{s}$ = 13 TeV}},
  \href{https://doi.org/10.1007/JHEP07(2023)073}{\emph{JHEP} {\bfseries 07}
  (2023) 073} [\href{https://arxiv.org/abs/2208.02717}{{\ttfamily
  2208.02717}}].

\bibitem{CMS:2023yay}
{\scshape CMS} collaboration, ``{Search for a standard model-like Higgs boson
  in the mass range between 70 and 110$~\mathrm{GeV}$ in the diphoton final
  state in proton-proton collisions at $\sqrt{s}=13~\mathrm{TeV}$}.''
  \url{https://cds.cern.ch/record/2852907/}, CERN, Geneva, 2023.

\bibitem{Arcangeletti}
C.~Arcangeletti, ``{ATLAS, LHC Seminar}.''
  {\url{https://indico.cern.ch/event/1281604/}}, 2023.

\bibitem{ATLAS:2023jzc}
{\scshape ATLAS} collaboration, ``{Search for diphoton resonances in the 66 to
  110 GeV mass range using 140 fb$^{-1}$ of 13 TeV $pp$ collisions collected
  with the ATLAS detector}.'' {\url{http://cds.cern.ch/record/2862024}}, CERN,
  Geneva, 2023.

\bibitem{CMS:2022arx}
{\scshape CMS} collaboration, ``{Search for dilepton resonances from decays of
  (pseudo)scalar bosons produced in association with a massive vector boson or
  top quark anti-top quark pair at $\sqrt{s}=13~\mathrm{TeV}$}.''
  {\url{http://cds.cern.ch/record/2815307}}, CERN, Geneva, 2022.

\bibitem{LEPWorkingGroupforHiggsbosonsearches}
{\scshape LEP Working Group for Higgs boson searches, ALEPH, DELPHI, L3, OPAL}
  collaboration, \emph{{Search for the standard model Higgs boson at LEP}},
  \href{https://doi.org/10.1016/S0370-2693(03)00614-2}{\emph{Phys. Lett. B}
  {\bfseries 565} (2003) 61}
  [\href{https://arxiv.org/abs/hep-ex/0306033}{{\ttfamily hep-ex/0306033}}].

\bibitem{ALEPH:2006tnd}
{\scshape ALEPH, DELPHI, L3, OPAL, LEP Working Group for Higgs Boson Searches}
  collaboration, \emph{{Search for neutral MSSM Higgs bosons at LEP}},
  \href{https://doi.org/10.1140/epjc/s2006-02569-7}{\emph{Eur. Phys. J. C}
  {\bfseries 47} (2006) 547}
  [\href{https://arxiv.org/abs/hep-ex/0602042}{{\ttfamily hep-ex/0602042}}].

\bibitem{Cao:2016uwt}
J.~Cao, X.~Guo, Y.~He, P.~Wu and Y.~Zhang, \emph{{Diphoton signal of the light
  Higgs boson in natural NMSSM}},
  \href{https://doi.org/10.1103/PhysRevD.95.116001}{\emph{Phys. Rev. D}
  {\bfseries 95} (2017) 116001}
  [\href{https://arxiv.org/abs/1612.08522}{{\ttfamily 1612.08522}}].

\bibitem{Heinemeyer:2021msz}
S.~Heinemeyer, C.~Li, F.~Lika, G.~Moortgat-Pick and S.~Paasch,
  \emph{{Phenomenology of a 96~GeV Higgs boson in the 2HDM with an additional
  singlet}}, \href{https://doi.org/10.1103/PhysRevD.106.075003}{\emph{Phys.
  Rev. D} {\bfseries 106} (2022) 075003}
  [\href{https://arxiv.org/abs/2112.11958}{{\ttfamily 2112.11958}}].

\bibitem{Biekotter:2021qbc}
T.~Biek\"otter, A.~Grohsjean, S.~Heinemeyer, C.~Schwanenberger and G.~Weiglein,
  \emph{{Possible indications for new Higgs bosons in the reach of the LHC:
  N2HDM and NMSSM interpretations}},
  \href{https://doi.org/10.1140/epjc/s10052-022-10099-1}{\emph{Eur. Phys. J. C}
  {\bfseries 82} (2022) 178}
  [\href{https://arxiv.org/abs/2109.01128}{{\ttfamily 2109.01128}}].

\bibitem{Biekotter:2019kde}
T.~Biek\"otter, M.~Chakraborti and S.~Heinemeyer, \emph{{A 96 GeV Higgs boson
  in the N2HDM}},
  \href{https://doi.org/10.1140/epjc/s10052-019-7561-2}{\emph{Eur. Phys. J. C}
  {\bfseries 80} (2020) 2} [\href{https://arxiv.org/abs/1903.11661}{{\ttfamily
  1903.11661}}].

\bibitem{Cao:2019ofo}
J.~Cao, X.~Jia, Y.~Yue, H.~Zhou and P.~Zhu, \emph{{96 GeV diphoton excess in
  seesaw extensions of the natural NMSSM}},
  \href{https://doi.org/10.1103/PhysRevD.101.055008}{\emph{Phys. Rev. D}
  {\bfseries 101} (2020) 055008}
  [\href{https://arxiv.org/abs/1908.07206}{{\ttfamily 1908.07206}}].

\bibitem{Biekotter:2022abc}
T.~Biek\"otter, S.~Heinemeyer and G.~Weiglein, \emph{{Excesses in the low-mass
  Higgs-boson search and the $W$-boson mass measurement}},
  \href{https://doi.org/10.1140/epjc/s10052-023-11635-3}{\emph{Eur. Phys. J. C}
  {\bfseries 83} (2023) 450}
  [\href{https://arxiv.org/abs/2204.05975}{{\ttfamily 2204.05975}}].

\bibitem{Iguro:2022dok}
S.~Iguro, T.~Kitahara and Y.~Omura, \emph{{Scrutinizing the 95\textendash{}100
  GeV di-tau excess in the top associated process}},
  \href{https://doi.org/10.1140/epjc/s10052-022-11028-y}{\emph{Eur. Phys. J. C}
  {\bfseries 82} (2022) 1053}
  [\href{https://arxiv.org/abs/2205.03187}{{\ttfamily 2205.03187}}].

\bibitem{Li:2022etb}
W.~Li, J.~Zhu, K.~Wang, S.~Ma, P.~Tian and H.~Qiao, \emph{{A light Higgs boson
  in the NMSSM confronted with the CMS di-photon and di-tau excesses}},
  \href{https://arxiv.org/abs/2212.11739}{{\ttfamily 2212.11739}}.

\bibitem{Cline:2019okt}
J.~M. Cline and T.~Toma, \emph{{Pseudo-Goldstone dark matter confronts cosmic
  ray and collider anomalies}},
  \href{https://doi.org/10.1103/PhysRevD.100.035023}{\emph{Phys. Rev. D}
  {\bfseries 100} (2019) 035023}
  [\href{https://arxiv.org/abs/1906.02175}{{\ttfamily 1906.02175}}].

\bibitem{Biekotter:2021ovi}
T.~Biek\"otter and M.~O. Olea-Romacho, \emph{{Reconciling Higgs physics and
  pseudo-Nambu-Goldstone dark matter in the S2HDM using a genetic algorithm}},
  \href{https://doi.org/10.1007/JHEP10(2021)215}{\emph{JHEP} {\bfseries 10}
  (2021) 215} [\href{https://arxiv.org/abs/2108.10864}{{\ttfamily
  2108.10864}}].

\bibitem{Crivellin:2017upt}
A.~Crivellin, J.~Heeck and D.~M\"uller, \emph{{Large $h\to b s$ in generic
  two-Higgs-doublet models}},
  \href{https://doi.org/10.1103/PhysRevD.97.035008}{\emph{Phys. Rev. D}
  {\bfseries 97} (2018) 035008}
  [\href{https://arxiv.org/abs/1710.04663}{{\ttfamily 1710.04663}}].

\bibitem{Cacciapaglia:2016tlr}
G.~Cacciapaglia, A.~Deandrea, S.~Gascon-Shotkin, S.~Le~Corre, M.~Lethuillier
  and J.~Tao, \emph{{Search for a lighter Higgs boson in Two Higgs Doublet
  Models}}, \href{https://doi.org/10.1007/JHEP12(2016)068}{\emph{JHEP}
  {\bfseries 12} (2016) 068}
  [\href{https://arxiv.org/abs/1607.08653}{{\ttfamily 1607.08653}}].

\bibitem{Abdelalim:2020xfk}
A.~A. Abdelalim, B.~Das, S.~Khalil and S.~Moretti, \emph{{Di-photon decay of a
  light Higgs state in the BLSSM}},
  \href{https://doi.org/10.1016/j.nuclphysb.2022.116013}{\emph{Nucl. Phys. B}
  {\bfseries 985} (2022) 116013}
  [\href{https://arxiv.org/abs/2012.04952}{{\ttfamily 2012.04952}}].

\bibitem{Biekotter:2022jyr}
T.~Biek\"otter, S.~Heinemeyer and G.~Weiglein, \emph{{Mounting evidence for a
  95 GeV Higgs boson}},
  \href{https://doi.org/10.1007/JHEP08(2022)201}{\emph{JHEP} {\bfseries 08}
  (2022) 201} [\href{https://arxiv.org/abs/2203.13180}{{\ttfamily
  2203.13180}}].

\bibitem{Biekotter:2023jld}
T.~Biek\"otter, S.~Heinemeyer and G.~Weiglein, \emph{{The CMS di-photon excess
  at 95 GeV in view of the LHC Run 2 results}},
  \href{https://doi.org/10.1016/j.physletb.2023.138217}{\emph{Phys. Lett. B}
  {\bfseries 846} (2023) 138217}
  [\href{https://arxiv.org/abs/2303.12018}{{\ttfamily 2303.12018}}].

\bibitem{Azevedo:2023zkg}
D.~Azevedo, T.~Biek\"otter and P.~M. Ferreira, \emph{{2HDM interpretations of
  the CMS diphoton excess at 95 GeV}},
  \href{https://doi.org/10.1007/JHEP11(2023)017}{\emph{JHEP} {\bfseries 11}
  (2023) 017} [\href{https://arxiv.org/abs/2305.19716}{{\ttfamily
  2305.19716}}].

\bibitem{Biekotter:2023oen}
T.~Biek\"otter, S.~Heinemeyer and G.~Weiglein, \emph{{95.4~GeV diphoton excess
  at ATLAS and CMS}},
  \href{https://doi.org/10.1103/PhysRevD.109.035005}{\emph{Phys. Rev. D}
  {\bfseries 109} (2024) 035005}
  [\href{https://arxiv.org/abs/2306.03889}{{\ttfamily 2306.03889}}].

\bibitem{Cao:2024axg}
J.~Cao, X.~Jia and J.~Lian, \emph{{Unified Interpretation of Muon g-2 anomaly,
  95 GeV Diphoton, and $b\bar{b}$ Excesses in the General Next-to-Minimal
  Supersymmetric Standard Model}},
  \href{https://arxiv.org/abs/2402.15847}{{\ttfamily 2402.15847}}.

\bibitem{Wang:2024bkg}
K.~Wang and J.~Zhu, \emph{{A 95 GeV light Higgs in the top-pair-associated
  diphoton channel at the LHC in the Minimal Dilaton Model}},
  \href{https://arxiv.org/abs/2402.11232}{{\ttfamily 2402.11232}}.

\bibitem{Li:2023kbf}
W.~Li, H.~Qiao, K.~Wang and J.~Zhu, \emph{{Light dark matter confronted with
  the 95 GeV diphoton excess}},
  \href{https://arxiv.org/abs/2312.17599}{{\ttfamily 2312.17599}}.

\bibitem{Dev:2023kzu}
P.~S.~B. Dev, R.~N. Mohapatra and Y.~Zhang, \emph{{Explanation of the 95 GeV
  \ensuremath{\gamma}\ensuremath{\gamma} and $b\bar{b}$ excesses in the minimal
  left-right symmetric model}},
  \href{https://doi.org/10.1016/j.physletb.2024.138481}{\emph{Phys. Lett. B}
  {\bfseries 849} (2024) 138481}
  [\href{https://arxiv.org/abs/2312.17733}{{\ttfamily 2312.17733}}].

\bibitem{Borah:2023hqw}
D.~Borah, S.~Mahapatra, P.~K. Paul and N.~Sahu, \emph{{Scotogenic
  U(1)L\ensuremath{\mu}-L\ensuremath{\tau} origin of (g-2)\ensuremath{\mu},
  W-mass anomaly and 95~GeV excess}},
  \href{https://doi.org/10.1103/PhysRevD.109.055021}{\emph{Phys. Rev. D}
  {\bfseries 109} (2024) 055021}
  [\href{https://arxiv.org/abs/2310.11953}{{\ttfamily 2310.11953}}].

\bibitem{Cao:2023gkc}
J.~Cao, X.~Jia, J.~Lian and L.~Meng, \emph{{95~GeV diphoton and $b\bar b$
  excesses in the general next-to-minimal supersymmetric standard model}},
  \href{https://doi.org/10.1103/PhysRevD.109.075001}{\emph{Phys. Rev. D}
  {\bfseries 109} (2024) 075001}
  [\href{https://arxiv.org/abs/2310.08436}{{\ttfamily 2310.08436}}].

\bibitem{Ellwanger:2023zjc}
U.~Ellwanger and C.~Hugonie, \emph{{Additional Higgs Bosons near 95 and 650 GeV
  in the NMSSM}},
  \href{https://doi.org/10.1140/epjc/s10052-023-12315-y}{\emph{Eur. Phys. J. C}
  {\bfseries 83} (2023) 1138}
  [\href{https://arxiv.org/abs/2309.07838}{{\ttfamily 2309.07838}}].

\bibitem{Aguilar-Saavedra:2023tql}
J.~A. Aguilar-Saavedra, H.~B. C\^amara, F.~R. Joaquim and J.~F. Seabra,
  \emph{{Confronting the 95 GeV excesses within the U(1)'-extended
  next-to-minimal 2HDM}},
  \href{https://doi.org/10.1103/PhysRevD.108.075020}{\emph{Phys. Rev. D}
  {\bfseries 108} (2023) 075020}
  [\href{https://arxiv.org/abs/2307.03768}{{\ttfamily 2307.03768}}].

\bibitem{Ashanujjaman:2023etj}
S.~Ashanujjaman, S.~Banik, G.~Coloretti, A.~Crivellin, B.~Mellado and A.-T.
  Mulaudzi, \emph{{SU(2)L triplet scalar as the origin of the 95~GeV excess?}},
  \href{https://doi.org/10.1103/PhysRevD.108.L091704}{\emph{Phys. Rev. D}
  {\bfseries 108} (2023) L091704}
  [\href{https://arxiv.org/abs/2306.15722}{{\ttfamily 2306.15722}}].

\bibitem{Bhattacharya:2023lmu}
S.~Bhattacharya, G.~Coloretti, A.~Crivellin, S.-E. Dahbi, Y.~Fang, M.~Kumar
  et~al., \emph{{Growing Excesses of New Scalars at the Electroweak Scale}},
  \href{https://arxiv.org/abs/2306.17209}{{\ttfamily 2306.17209}}.

\bibitem{Dutta:2023cig}
J.~Dutta, J.~Lahiri, C.~Li, G.~Moortgat-Pick, S.~F. Tabira and J.~A. Ziegler,
  \emph{{Dark Matter Phenomenology in 2HDMS in light of the 95 GeV excess}},
  \href{https://arxiv.org/abs/2308.05653}{{\ttfamily 2308.05653}}.

\bibitem{Ellwanger:2024txc}
U.~Ellwanger and C.~Hugonie, \emph{{NMSSM with correct relic density and an
  additional 95 GeV Higgs boson}},
  \href{https://arxiv.org/abs/2403.16884}{{\ttfamily 2403.16884}}.

\bibitem{Diaz:2024yfu}
M.~A. Diaz, G.~Cerro, S.~Dasmahapatra and S.~Moretti, \emph{{Bayesian Active
  Search on Parameter Space: a 95 GeV Spin-0 Resonance in the ($B-L$)SSM}},
  \href{https://arxiv.org/abs/2404.18653}{{\ttfamily 2404.18653}}.

\bibitem{Ellwanger:2024vvs}
U.~Ellwanger, C.~Hugonie, S.~F. King and S.~Moretti, \emph{{NMSSM Explanation
  for Excesses in the Search for Neutralinos and Charginos and a 95 GeV Higgs
  Boson}},  \href{https://arxiv.org/abs/2404.19338}{{\ttfamily 2404.19338}}.

\bibitem{Ayazi:2024fmn}
S.~Y. Ayazi, M.~Hosseini, S.~Paktinat~Mehdiabadi and R.~Rouzbehi, \emph{{The
  Vector Dark Matter, LHC Constraints Including a 95 GeV Light Higgs Boson}},
  \href{https://arxiv.org/abs/2405.01132}{{\ttfamily 2405.01132}}.

\bibitem{Arhrib:2024wjj}
A.~Arhrib, K.~H. Phan, V.~Q. Tran and T.-C. Yuan, \emph{{When Standard Model
  Higgs Meets Its Lighter 95 GeV Higgs}},
  \href{https://arxiv.org/abs/2405.03127}{{\ttfamily 2405.03127}}.

\bibitem{Gao:2024qag}
J.~Gao, J.~Ma, L.~Wang and H.~Xu, \emph{{A 95 GeV Higgs boson and spontaneous
  CP-violation at the finite temperature}},
  \href{https://arxiv.org/abs/2408.03705}{{\ttfamily 2408.03705}}.

\bibitem{Benbrik:2022azi}
R.~Benbrik, M.~Boukidi, S.~Moretti and S.~Semlali, \emph{{Explaining the 96 GeV
  Di-photon anomaly in a generic 2HDM Type-III}},
  \href{https://doi.org/10.1016/j.physletb.2022.137245}{\emph{Phys. Lett. B}
  {\bfseries 832} (2022) 137245}
  [\href{https://arxiv.org/abs/2204.07470}{{\ttfamily 2204.07470}}].

\bibitem{Benbrik:2022tlg}
R.~Benbrik, M.~Boukidi, S.~Moretti and S.~Semlali, \emph{{Probing a 96 GeV
  Higgs Boson in the Di-Photon Channel at the LHC}},
  \href{https://doi.org/10.22323/1.414.0547}{\emph{PoS} {\bfseries ICHEP2022}
  (2022) 547} [\href{https://arxiv.org/abs/2211.11140}{{\ttfamily
  2211.11140}}].

\bibitem{Belyaev:2023xnv}
A.~Belyaev, R.~Benbrik, M.~Boukidi, M.~Chakraborti, S.~Moretti and S.~Semlali,
  \emph{{Explanation of the Hints for a 95 GeV Higgs Boson within a 2-Higgs
  Doublet Model}},  \href{https://arxiv.org/abs/2306.09029}{{\ttfamily
  2306.09029}}.

\bibitem{Belyaev:2024lah}
A.~Belyaev, R.~Benbrik, M.~Boukidi, M.~Chakraborti, S.~Moretti and S.~Semlali,
  \emph{{Probing 95 GeV Higgs in the 2HDM Type-III}},
  \href{https://doi.org/10.22323/1.450.0299}{\emph{PoS} {\bfseries LHCP2023}
  (2024) 299} [\href{https://arxiv.org/abs/2402.03998}{{\ttfamily
  2402.03998}}].

\bibitem{Benbrik:2024ptw}
R.~Benbrik, M.~Boukidi and S.~Moretti, \emph{{Superposition of CP-Even and
  CP-Odd Higgs Resonances: Explaining the 95 GeV Excesses within a Two-Higgs
  Doublet Model}},  \href{https://arxiv.org/abs/2405.02899}{{\ttfamily
  2405.02899}}.

\bibitem{Khanna:2024bah}
A.~Khanna, S.~Moretti and A.~Sarkar, \emph{{Explaining 95 (or so) GeV Anomalies
  in the 2-Higgs Doublet Model Type-I}},
  \href{https://arxiv.org/abs/2409.02587}{{\ttfamily 2409.02587}}.

\bibitem{Bower:2002zx}
G.~R. Bower, T.~Pierzchala, Z.~Was and M.~Worek, \emph{{Measuring the Higgs
  boson's parity using tau ---\ensuremath{>} rho nu}},
  \href{https://doi.org/10.1016/S0370-2693(02)02445-0}{\emph{Phys. Lett. B}
  {\bfseries 543} (2002) 227}
  [\href{https://arxiv.org/abs/hep-ph/0204292}{{\ttfamily hep-ph/0204292}}].

\bibitem{Desch:2003rw}
K.~Desch, A.~Imhof, Z.~Was and M.~Worek, \emph{{Probing the CP nature of the
  Higgs boson at linear colliders with tau spin correlations: The Case of mixed
  scalar - pseudoscalar couplings}},
  \href{https://doi.org/10.1016/j.physletb.2003.10.074}{\emph{Phys. Lett. B}
  {\bfseries 579} (2004) 157}
  [\href{https://arxiv.org/abs/hep-ph/0307331}{{\ttfamily hep-ph/0307331}}].

\bibitem{Berge:2013jra}
S.~Berge, W.~Bernreuther and H.~Spiesberger, \emph{{Higgs CP properties using
  the $\tau$ decay modes at the ILC}},
  \href{https://doi.org/10.1016/j.physletb.2013.11.006}{\emph{Phys. Lett. B}
  {\bfseries 727} (2013) 488}
  [\href{https://arxiv.org/abs/1308.2674}{{\ttfamily 1308.2674}}].

\bibitem{Berge:2008wi}
S.~Berge, W.~Bernreuther and J.~Ziethe, \emph{{Determining the CP parity of
  Higgs bosons at the LHC in their tau decay channels}},
  \href{https://doi.org/10.1103/PhysRevLett.100.171605}{\emph{Phys. Rev. Lett.}
  {\bfseries 100} (2008) 171605}
  [\href{https://arxiv.org/abs/0801.2297}{{\ttfamily 0801.2297}}].

\bibitem{Berge:2008dr}
S.~Berge and W.~Bernreuther, \emph{{Determining the CP parity of Higgs bosons
  at the LHC in the tau to 1-prong decay channels}},
  \href{https://doi.org/10.1016/j.physletb.2008.12.065}{\emph{Phys. Lett. B}
  {\bfseries 671} (2009) 470}
  [\href{https://arxiv.org/abs/0812.1910}{{\ttfamily 0812.1910}}].

\bibitem{Berge:2011ij}
S.~Berge, W.~Bernreuther, B.~Niepelt and H.~Spiesberger, \emph{{How to pin down
  the CP quantum numbers of a Higgs boson in its tau decays at the LHC}},
  \href{https://doi.org/10.1103/PhysRevD.84.116003}{\emph{Phys. Rev. D}
  {\bfseries 84} (2011) 116003}
  [\href{https://arxiv.org/abs/1108.0670}{{\ttfamily 1108.0670}}].

\bibitem{Berge:2014sra}
S.~Berge, W.~Bernreuther and S.~Kirchner, \emph{{Determination of the Higgs
  CP-mixing angle in the tau decay channels at the LHC including the
  Drell\textendash{}Yan background}},
  \href{https://doi.org/10.1140/epjc/s10052-014-3164-0}{\emph{Eur. Phys. J. C}
  {\bfseries 74} (2014) 3164}
  [\href{https://arxiv.org/abs/1408.0798}{{\ttfamily 1408.0798}}].

\bibitem{Berge:2014wta}
S.~Berge, W.~Bernreuther and S.~Kirchner, \emph{{Determination of the Higgs
  CP-mixing angle in the tau decay channels}},
  \href{https://doi.org/10.1016/j.nuclphysbps.2015.09.129}{\emph{Nucl. Part.
  Phys. Proc.} {\bfseries 273-275} (2016) 841}
  [\href{https://arxiv.org/abs/1410.6362}{{\ttfamily 1410.6362}}].

\bibitem{Berge:2015nua}
S.~Berge, W.~Bernreuther and S.~Kirchner, \emph{{Prospects of constraining the
  Higgs boson\textquoteright{}s CP nature in the tau decay channel at the
  LHC}}, \href{https://doi.org/10.1103/PhysRevD.92.096012}{\emph{Phys. Rev. D}
  {\bfseries 92} (2015) 096012}
  [\href{https://arxiv.org/abs/1510.03850}{{\ttfamily 1510.03850}}].

\bibitem{Han:2016bvf}
T.~Han, S.~Mukhopadhyay, B.~Mukhopadhyaya and Y.~Wu, \emph{{Measuring the CP
  property of Higgs coupling to tau leptons in the VBF channel at the LHC}},
  \href{https://doi.org/10.1007/JHEP05(2017)128}{\emph{JHEP} {\bfseries 05}
  (2017) 128} [\href{https://arxiv.org/abs/1612.00413}{{\ttfamily
  1612.00413}}].

\bibitem{Antusch:2020ngh}
S.~Antusch, O.~Fischer, A.~Hammad and C.~Scherb, \emph{{Testing CP Properties
  of Extra Higgs States at the HL-LHC}},
  \href{https://doi.org/10.1007/JHEP03(2021)200}{\emph{JHEP} {\bfseries 03}
  (2021) 200} [\href{https://arxiv.org/abs/2011.10388}{{\ttfamily
  2011.10388}}].

\bibitem{Dutta:2021del}
S.~Dutta, A.~Goyal, M.~Kumar and A.~K. Swain, \emph{{Measuring CP nature of
  $h\tau {{\bar{\tau }}}$ coupling at $e^-p$ collider}},
  \href{https://doi.org/10.1140/epjc/s10052-022-10445-3}{\emph{Eur. Phys. J. C}
  {\bfseries 82} (2022) 530}
  [\href{https://arxiv.org/abs/2109.00329}{{\ttfamily 2109.00329}}].

\bibitem{Esmail:2024jdg}
W.~Esmail, A.~Hammad, M.~Nojiri and C.~Scherb, \emph{{Testing CP properties of
  the Higgs boson coupling to $\tau$ leptons with heterogeneous graphs}},
  \href{https://arxiv.org/abs/2409.06132}{{\ttfamily 2409.06132}}.

\bibitem{CMS-PAS-HIG-21-001}
{\scshape CMS} collaboration, \emph{{Searches for additional Higgs bosons and
  vector leptoquarks in $\tau\tau$ final states in proton-proton collisions at
  $\sqrt{s}=13~\mathrm{TeV}$}},  tech. rep., CERN, Geneva, 2022.
\newblock https://cds.cern.ch/record/2803739.

\bibitem{LEPWorkingGroupforHiggsbosonsearches:2003ing}
{\scshape LEP Working Group for Higgs boson searches, ALEPH, DELPHI, L3, OPAL}
  collaboration, \emph{{Search for the standard model Higgs boson at LEP}},
  \href{https://doi.org/10.1016/S0370-2693(03)00614-2}{\emph{Phys. Lett. B}
  {\bfseries 565} (2003) 61}
  [\href{https://arxiv.org/abs/hep-ex/0306033}{{\ttfamily hep-ex/0306033}}].

\bibitem{Hansen:2020ecn}
M.~C. Hansen, \emph{{Studies into measuring the Higgs CP-state in $H$ $\to$
  $\tau\tau$ decays at ATLAS}}, Ph.D. thesis, Bonn U., 2020.

\bibitem{Gianotti:2002xx}
F.~Gianotti et~al., \emph{{Physics potential and experimental challenges of the
  LHC luminosity upgrade}},
  \href{https://doi.org/10.1140/epjc/s2004-02061-6}{\emph{Eur. Phys. J. C}
  {\bfseries 39} (2005) 293}
  [\href{https://arxiv.org/abs/hep-ph/0204087}{{\ttfamily hep-ph/0204087}}].

\bibitem{Apollinari:2015wtw}
G.~Apollinari, O.~Br\"uning, T.~Nakamoto and L.~Rossi, \emph{{High Luminosity
  Large Hadron Collider HL-LHC}},
  \href{https://doi.org/10.5170/CERN-2015-005.1}{\emph{CERN Yellow Rep.} (2015)
  1} [\href{https://arxiv.org/abs/1705.08830}{{\ttfamily 1705.08830}}].

\bibitem{ATLAS:2022akr}
{\scshape ATLAS} collaboration, \emph{{Measurement of the CP properties of
  Higgs boson interactions with $\tau $-leptons with the ATLAS detector}},
  \href{https://doi.org/10.1140/epjc/s10052-023-11583-y}{\emph{Eur. Phys. J. C}
  {\bfseries 83} (2023) 563}
  [\href{https://arxiv.org/abs/2212.05833}{{\ttfamily 2212.05833}}].

\bibitem{CMS:2021sdq}
{\scshape CMS} collaboration, \emph{{Analysis of the $CP$ structure of the
  Yukawa coupling between the Higgs boson and $\tau$ leptons in proton-proton
  collisions at $ \sqrt{s} $ = 13 TeV}},
  \href{https://doi.org/10.1007/JHEP06(2022)012}{\emph{JHEP} {\bfseries 06}
  (2022) 012} [\href{https://arxiv.org/abs/2110.04836}{{\ttfamily
  2110.04836}}].

\bibitem{Christensen:2008py}
N.~D. Christensen and C.~Duhr, \emph{{FeynRules - Feynman rules made easy}},
  \href{https://doi.org/10.1016/j.cpc.2009.02.018}{\emph{Comput. Phys. Commun.}
  {\bfseries 180} (2009) 1614}
  [\href{https://arxiv.org/abs/0806.4194}{{\ttfamily 0806.4194}}].

\bibitem{Alloul:2013bka}
A.~Alloul, N.~D. Christensen, C.~Degrande, C.~Duhr and B.~Fuks,
  \emph{{FeynRules 2.0 - A complete toolbox for tree-level phenomenology}},
  \href{https://doi.org/10.1016/j.cpc.2014.04.012}{\emph{Comput. Phys. Commun.}
  {\bfseries 185} (2014) 2250}
  [\href{https://arxiv.org/abs/1310.1921}{{\ttfamily 1310.1921}}].

\bibitem{Harlander:2012pb}
R.~V. Harlander, S.~Liebler and H.~Mantler, \emph{{SusHi: A program for the
  calculation of Higgs production in gluon fusion and bottom-quark annihilation
  in the Standard Model and the MSSM}},
  \href{https://doi.org/10.1016/j.cpc.2013.02.006}{\emph{Comput. Phys. Commun.}
  {\bfseries 184} (2013) 1605}
  [\href{https://arxiv.org/abs/1212.3249}{{\ttfamily 1212.3249}}].

\bibitem{Harlander:2016hcx}
R.~V. Harlander, S.~Liebler and H.~Mantler, \emph{{SusHi Bento: Beyond NNLO and
  the heavy-top limit}},
  \href{https://doi.org/10.1016/j.cpc.2016.10.015}{\emph{Comput. Phys. Commun.}
  {\bfseries 212} (2017) 239}
  [\href{https://arxiv.org/abs/1605.03190}{{\ttfamily 1605.03190}}].

\bibitem{Choi:2021nql}
S.~Y. Choi, J.~S. Lee and J.~Park, \emph{{Decays of Higgs bosons in the
  Standard Model and beyond}},
  \href{https://doi.org/10.1016/j.ppnp.2021.103880}{\emph{Prog. Part. Nucl.
  Phys.} {\bfseries 120} (2021) 103880}
  [\href{https://arxiv.org/abs/2101.12435}{{\ttfamily 2101.12435}}].

\bibitem{Kramer:1993jn}
M.~Kramer, J.~H. Kuhn, M.~L. Stong and P.~M. Zerwas, \emph{{Prospects of
  measuring the parity of Higgs particles}},
  \href{https://doi.org/10.1007/BF01557231}{\emph{Z. Phys. C} {\bfseries 64}
  (1994) 21} [\href{https://arxiv.org/abs/hep-ph/9404280}{{\ttfamily
  hep-ph/9404280}}].

\bibitem{Tsai:1971vv}
Y.-S. Tsai, \emph{{Decay Correlations of Heavy Leptons in e+ e-
  ---\ensuremath{>} Lepton+ Lepton-}},
  \href{https://doi.org/10.1103/PhysRevD.13.771}{\emph{Phys. Rev. D} {\bfseries
  4} (1971) 2821}.

\bibitem{Alwall:2011uj}
J.~Alwall, M.~Herquet, F.~Maltoni, O.~Mattelaer and T.~Stelzer, \emph{{MadGraph
  5 : Going Beyond}},
  \href{https://doi.org/10.1007/JHEP06(2011)128}{\emph{JHEP} {\bfseries 06}
  (2011) 128} [\href{https://arxiv.org/abs/1106.0522}{{\ttfamily 1106.0522}}].

\bibitem{Alwall:2014hca}
J.~Alwall, R.~Frederix, S.~Frixione, V.~Hirschi, F.~Maltoni, O.~Mattelaer
  et~al., \emph{{The automated computation of tree-level and next-to-leading
  order differential cross sections, and their matching to parton shower
  simulations}}, \href{https://doi.org/10.1007/JHEP07(2014)079}{\emph{JHEP}
  {\bfseries 07} (2014) 079} [\href{https://arxiv.org/abs/1405.0301}{{\ttfamily
  1405.0301}}].

\bibitem{Hagiwara:2012vz}
K.~Hagiwara, T.~Li, K.~Mawatari and J.~Nakamura, \emph{{TauDecay: a library to
  simulate polarized tau decays via FeynRules and MadGraph5}},
  \href{https://doi.org/10.1140/epjc/s10052-013-2489-4}{\emph{Eur. Phys. J. C}
  {\bfseries 73} (2013) 2489}
  [\href{https://arxiv.org/abs/1212.6247}{{\ttfamily 1212.6247}}].

\bibitem{Sjostrand:2014zea}
T.~Sjöstrand, S.~Ask, J.~R. Christiansen, R.~Corke, N.~Desai, P.~Ilten et~al.,
  \emph{{An Introduction to PYTHIA 8.2}},
  \href{https://doi.org/10.1016/j.cpc.2015.01.024}{\emph{Comput. Phys. Commun.}
  {\bfseries 191} (2015) 159}
  [\href{https://arxiv.org/abs/1410.3012}{{\ttfamily 1410.3012}}].

\bibitem{Bierlich:2022pfr}
C.~Bierlich et~al., \emph{{A comprehensive guide to the physics and usage of
  PYTHIA 8.3}},
  \href{https://doi.org/10.21468/SciPostPhysCodeb.8}{\emph{SciPost Phys.
  Codeb.} {\bfseries 2022} (2022) 8}
  [\href{https://arxiv.org/abs/2203.11601}{{\ttfamily 2203.11601}}].

\bibitem{deFavereau:2013fsa}
{\scshape DELPHES 3} collaboration, \emph{{DELPHES 3, A modular framework for
  fast simulation of a generic collider experiment}},
  \href{https://doi.org/10.1007/JHEP02(2014)057}{\emph{JHEP} {\bfseries 02}
  (2014) 057} [\href{https://arxiv.org/abs/1307.6346}{{\ttfamily 1307.6346}}].

\bibitem{Cacciari:2008gp}
M.~Cacciari, G.~P. Salam and G.~Soyez, \emph{{The Anti-k(t) jet clustering
  algorithm}}, \href{https://doi.org/10.1088/1126-6708/2008/04/063}{\emph{JHEP}
  {\bfseries 04} (2008) 063} [\href{https://arxiv.org/abs/0802.1189}{{\ttfamily
  0802.1189}}].

\bibitem{Maguire:2017ypu}
E.~Maguire, L.~Heinrich and G.~Watt, \emph{{HEPData: a repository for high
  energy physics data}},
  \href{https://doi.org/10.1088/1742-6596/898/10/102006}{\emph{J. Phys. Conf.
  Ser.} {\bfseries 898} (2017) 102006}
  [\href{https://arxiv.org/abs/1704.05473}{{\ttfamily 1704.05473}}].

\bibitem{hepdata.128147}
{CMS Collaboration}, ``Searches for additional higgs bosons and for vector
  leptoquarks in $\tau\tau$ final states in proton-proton collisions at
  $\sqrt{s}$ = 13 tev.'' \url{https://www.hepdata.net/record/ins2132368}, 2022.
\newblock \url{https://doi.org/10.17182/hepdata.128147}.

\end{thebibliography}\endgroup
%***********************************
%***********************************
\end{document}